\documentclass[twocolumn,english,aps,prb,floats]{revtex4-2}
\usepackage[T1]{fontenc}
\usepackage[utf8]{inputenc}
\usepackage{babel}
\usepackage{bm}
\usepackage{amsmath}
\usepackage{amssymb}
\usepackage{graphicx}
\usepackage{esint}
\usepackage[pdfusetitle,
 bookmarks=true,bookmarksnumbered=false,bookmarksopen=false,
 breaklinks=false,pdfborder={0 0 1},backref=false,colorlinks=false]
 {hyperref}
\begin{document}
\title{Massive dynamics of skyrmions in ferrimagnetic films}
\author{Dmitry A. Garanin and Eugene M. Chudnovsky}
\affiliation{Physics Department, Herbert H. Lehman College and Graduate School,
The City University of New York, 250 Bedford Park Boulevard West,
Bronx, New York 10468-1589, USA }
\date{\today}
\begin{abstract}
Deformations of skyrmions arising from the presence of more than one
magnetic sublattice lead to their massive dynamics in ferrimagnets
as compared to the massless dynamics in 2D ferromagnets. This results
in the gyroscopic motion of skyrmions, which manifests as skyrmion
cyclotron resonance that can be excited by microwaves or spin currents.
We investigate analytically and numerically the motion and resonant
oscillations of individual skyrmions and skyrmion lattices in the
presence of dissipation in a two-sublattice transition-metal -- rare-earth
(TM/RE) system. The focus is on the dependence of the skyrmion dynamics
on the RE concentration. Parameters of the CoGd ferrimagnet are utilized
in the numerical work. The massive dynamics of skyrmions in ferrimagnets,
as well as the spectrum of their excitations, undergo significant
changes near the angular momentum compensation point, which should
not be difficult to detect in experiments. 
\end{abstract}
\maketitle

\section{Introduction}

Frequencies of excitation modes in ferrimagnets have been studied
for over 70 years, starting with the comprehensive investigation by
Wangsness \citep{Wangsness-PR1953} and the work of Kittel \citep{Kittel-PR1959}
on ferromagnetic resonance in rare-earth garnets. Later on, several
authors re-derived their results with a focus on the angular momentum
compensation point (AMC) where excitation frequencies go up \citep{Lin-PRB1988,Zhang-JPhys1997,Karchev-JPhys2008,Okuno-APLExpress2019,Haltz-PRB2022,Ivanov-PRB2023,Sanchez-PRB2025,DG-EC-PRB2026}.
The latter was confirmed by numerous experiments \citep{Pardavi-JMMM2000,Binder-PRB2006,Stanciu-PRB2006,Okuno-APLExpress2019,Kim-Nature2020}
and prompted the increased interest to the potential of ferrimagnets
for applications. Similar to antiferromagnets, due to the hardening
of spin excitations at the AMC, ferrimagnets permit fast dynamics
required for the information processing \citep{Binder-PRB2006,Arena-PRApplied2017,Siddiqui-PRB2018,Ivanov-review,Bonfiglio-PRB2019,Kim-Nature2020,DavydovaJOP2020,Yurlov-PRB2021,Joo-materials2021,Chanda-PRB2021,Kim-NatMat2022,Guo-PRB2022,Zhang-PRB2022,Chen-Materials2025,Moreno-PRB2025,Ciccarelli-AP2025}.
What makes them more suited for that purpose than antiferromagnets
is the possession of uncompensated magnetic moments at the AMC due
to the different gyromagnetic ratios of atoms belonging to different
magnetic sublattices.

Within a certain range of parameters and magnetic fields, films of
many ferrimagnetic materials possess topological defects -- skyrmions.
The interest in skyrmions, besides their mathematical beauty, has
been inspired by their potential for topologically protected information
technology \citep{APL-2021}. On changing the magnetic field, skyrmion
lattices (SkL) often replace conventional magnetic domain structures.
Excitation modes of individual skyrmions and SkL have been studied
in ferromagnets \citep{Mochizuki-PRL2012,Onose-PRL2012,DA-RJ-EC-PRB2020,Aqeel-PRL2021,Satywali-NatCom2021,Lee-JPhys2022,Li-JPhys2023,DG-EC-PRB2025},
but very scarcely, if at all, in ferrimagnets. In this connection,
it is important to note that the dynamics of skyrmions depends on
whether they are massive or massless.

The mass of a magnetic skyrmion has been a puzzle of skyrmion physics
\citep{Makhfudz-PRL2012,Buttner-Nature2015,Shiino2017,Lin-PRB2017,Psaroudaki-PRX2017,Kravchuk-PRB2018,Li-PRB2018}.
While the finite mass of a ferromagnetic domain wall has been computed
\citep{Doring} and subsequently measured by various methods, the
inertial mass associated with the motion of the center of the topological
charge of a skyrmion in a uniform 2D film has been rigorously shown
to be exactly zero \citep{Komineas2015,EC-DA-EPL2025} (see also Ref.
\citep{Tchernyshyov}). It was demonstrated theoretically that a nonzero
skyrmion mass can arise from its confinement in a nanoring \citep{Liu-MMM2020}
or nanotrack \citep{EC-DA-EPL2025}, or it can be generated by the
interaction with other degrees of freedom, such as phonons \citep{Capic-PRB2020}.
Experiments on the skyrmion mass have been inconclusive so far, with
various authors citing numbers from as small as $10^{-27}$kg to as
large as $10^{-22}$kg.

\begin{figure}
\centering{}\includegraphics[width=8cm]{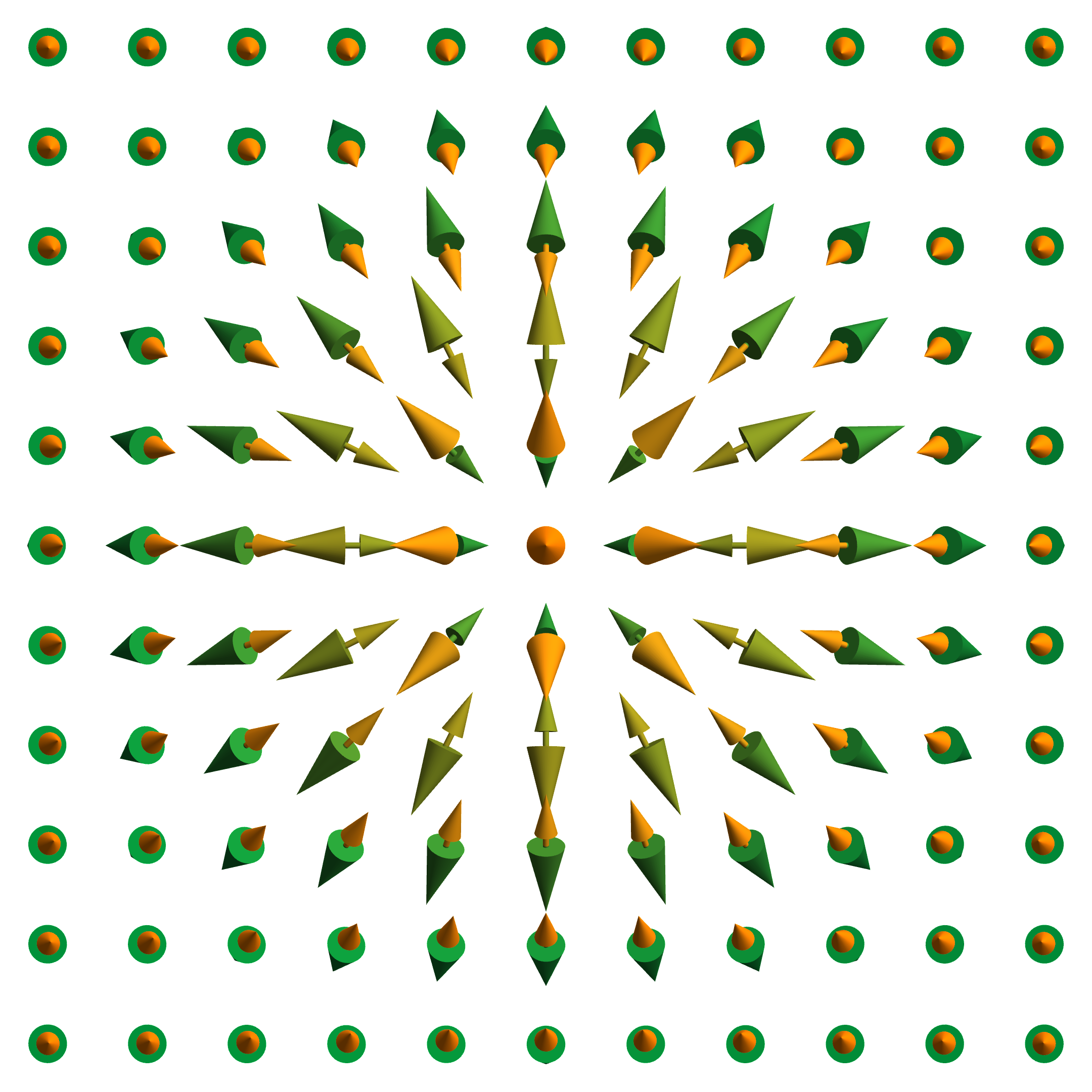}
\caption{Two-sublattice skyrmion in a TM/RE ferrimagnet. Far from the skyrmion's
core, TM spins (larger cones) are directed down (green), and RE spins
(smaller cones) are directed up (orange). }\label{Fig_Ferrimagnetic_skyrmion}
\end{figure}

The question of the skyrmion mass acquires special importance when
addressing the excitation spectrum of a ferrimagnet in the presence
of skyrmions. The spin texture of a ferrimagnetic skyrmion is illustrated
in Fig.\ \ref{Fig_Ferrimagnetic_skyrmion} for a transition metal
(TM) -- rare earth (RE) ferrimagnet having antiferromagnetic exchange
interaction between the TM and RE sublattices. A specific deformation
of the skyrmion corresponding to the tiny splitting of its spin texture
into two shifted skyrmions belonging to different spin sublattices
(shown by different colors in Fig. \ref{Fig_Ferrimagnetic_skyrmion})
has been shown to generate skyrmion inertia \citep{Panigrahy2022,Nowak-PRB2023,Lau2025,EC-DG-arXiv}.
This results in the mass term in the Thiele equation \citep{Thiele}
describing skyrmion dynamics. In this paper, we derive the Thiele
equation for a ferrimagnet, taking into account the above mentioned
deformations of the skyrmion and the dissipation of the skyrmion motion.
The skyrmion mass has a universal form determined by the intersublattice
exchange interaction, and is independent of the spin densities of
the sublattices. In a ferrimagnetic film of finite thickness, it grows
linearly with the number of atomic layers, with a typical value for
one atomic layer in a TM/RE ferrimagnet being of the order of the
proton mass.

One of the consequences of the finite skyrmion mass is its gyroscopic
motion, resembling the circular motion of the electron in the magnetic
field. It can be excited for individual skyrmions by the spin currents
or microwaves, similar to how electron cyclotron resonance is excited
in metals \citep{ECR}. For a Belavin-Polyakov (BP) pure-exchange
skyrmion \citep{BP} the frequency of the skyrmion cyclotron resonance
(SCR) in a two-sublattice ferrimagnet, that follows from our analytical
model, is $\Omega=(J'/\hbar)|S-c\Sigma|$, where $J'$ is the intersublattice
exchange interaction between TM atoms of spin $S$ and RE atoms of
spin $\Sigma>S$, and $c$ is the ratio of their concentrations. Numerically,
we find that this simple formula for the SCR frequency possesses high
universality. It works remarkably well for individual skyrmions and
skyrmion lattices when interactions other than the nearest-neighbor
exchange, such as the magnetic anisotropy and the Dzyaloshinskii-Moriya
interaction (DMI), are added to the Hamiltonian.

Our model is based on a spin-lattice Hamiltonian tailored for a CoGd
ferrimagnet. The numerical methods include the energy minimization
and the computation of the dynamical evolution of the system described
by the Landau-Lifshitz equation for a many-spin system on the lattice.
While in the analytical approach, the dissipation is introduced phenomenologically,
in the numerical approach, it appears naturally due to the nonlinearity
of the spin dynamics. We obtain the frequencies of spin excitations
by computing the fluctuation spectrum (FS) of the system. Contrary
to the uniform ferrimagnetic modes whose frequencies in the absence
of skyrmions go up in the region of intermediate RE concentrations
($c\sim S/\Sigma$) \citep{DG-EC-PRB2026}, the SCR frequency $\Omega=(J'/\hbar)|S-c\Sigma|$
goes down and vanishes at the AMC point. In this region, in the presence
of the DMI and the anisotropy, as in CoGd, the SCR mode strongly hybridizes
with other ferrimagnetic modes, pushing them down. Such a dramatic
change in the behavior of the ferrimagnetic modes in the presence
of skyrmions should not be difficult to observe in experiments.

The paper is structured as follows. The model and excitation spectrum
of a TM/RE ferrimagnet without topological defects are discussed in
Sec. \ref{Sec_The-model}. Coupled dissipative Thiele equations for
a ferrimagnetic skyrmion, which permit separation of skyrmion centers
in the two sublattices, are presented in Sec. \ref{Sec_Thiele-equations}.
The massive Thiele equation describing the dynamics of the twin skyrmion,
its effective mass, and the frequency of the cyclotron resonance are
obtained in Sec. \ref{Sec_Skyrmion-mass}. Section \ref{Sec_Analysis}
is devoted to the analytical solution of the obtained massive Thiele
equation. The numerical methods are introduced in Sec. \ref{Sec_Numerical-methods}.
The numerical results on skyrmion trajectories and the dependence
of the excitation spectrum of SkL on the RE concentration are presented
in Sec. \ref{Sec_Numerical-results}.

\section{The model}

\label{Sec_The-model}

Keeping ferrimagnetic films in mind, we consider a square lattice
model of a TM/RE ferrimagnet with skyrmions, described by the Hamiltonian
\begin{eqnarray}
{\cal H} & = & -\frac{1}{2}\sum_{ij}J_{ij}{\bf S}_{i}\cdot{\bf S}_{j}+J'\sum_{i}{\bf S}_{i}\cdot p_{i}\bm{\Sigma}_{i}\nonumber \\
 &  & -\left[\mathbf{H}+\mathbf{h}(t)\right]\cdot\sum_{i}\left(g\mu_{B}\mathbf{S}_{i}+g'\mu_{B}p_{i}\boldsymbol{\Sigma}_{i}\right)-\frac{D}{2}\sum_{i}S_{i,z}^{2}\nonumber \\
 & + & A\sum_{i}\left[({\bf S}_{i}\times{\bf S}_{i+\delta_{x}})_{x}+{\bf S}_{i}\times{\bf S}_{i+\delta_{y}})_{y}\right].\label{Ham}
\end{eqnarray}
Here $J_{ij}$ is the ferromagnetic nearest-neighbor exchange within
the square TM-sublattice of spins ${\bf S}_{i}$ with the coupling
constant $J>0$ and lattice spacing $a$. The RE spins $\bm{\Sigma}_{i}$
occupy a similar square lattice and are coupled to the TM spins with
the coupling constant $J'>0$. These spins are diluted, which is taken
into account by the occupation factors $p_{i}=0,1$, so that $\left\langle p_{i}\right\rangle =c$.
As a variant, we also consider the \textit{no-disorder} model, in
which all RE spins have the same length $c\Sigma$. Other terms are
the easy-axis anisotropy and the DMI of the Bloch type within the
TM sublattice. This kind of DMI favors skyrmions with the counterclockwise
orientation of the in-plane spin components for $A>0$. The Néel-type
DMI always yields identical results, so it will not be considered
here. Having CoGd in mind, we chose $S=3/2$, $\Sigma=7/2$, $J'/J=0.2$,
$D/J=0.03$, $g=2.2$, $g'=2$, and $A/J=0.1$. In the Zeeman term,
$\mathbf{H}$ is the static applied magnetic field and $\mathbf{h}(t)$
is the microwave (MW) field. 

The continuous version of this lattice model has the energy density
given by
\begin{equation}
\epsilon=\epsilon_{\mathrm{ex}}+\epsilon_{Z}+\epsilon_{A}+\epsilon_{\mathrm{DMI}}+\ldots,
\end{equation}
where
\begin{eqnarray}
\epsilon_{\mathrm{ex}} & = & a^{d+2}\frac{J}{2}\bm{\nabla}\rho_{S,\alpha}\cdot\bm{\nabla}\rho_{S,\alpha}+a^{d}J'\boldsymbol{\rho}_{S}\cdot\boldsymbol{\rho}_{\Sigma}\nonumber \\
\epsilon_{Z} & = & -\mathbf{H}\cdot\left(g\mu_{B}\boldsymbol{\rho}_{S}+g'\mu_{B}\boldsymbol{\rho}_{\Sigma}\right)\nonumber \\
\epsilon_{A} & = & -a^{d}\frac{D}{2}\rho_{S,z}^{2}\nonumber \\
\epsilon_{\mathrm{DMI}} & = & -a^{d+1}A\boldsymbol{\rho}_{S}\cdot\left(\nabla\times\boldsymbol{\rho}_{S}\right),\label{Energy_density}
\end{eqnarray}
where $d$ is the hypercubic lattice dimension (here $d=2$), $\boldsymbol{\rho}_{S}=\mathbf{S}/a^{d}$
and $\boldsymbol{\rho}_{\Sigma}=\boldsymbol{\Sigma}/a^{d}$ are TM
and RE spin densities, the summation over repeated indices is implied
in the exchange energy $\epsilon_{\mathrm{ex}}$, and in $\epsilon_{\mathrm{DMI}}$
the derivatives over $z$ should be discarded. 

The dynamics of the lattice spins is described by the Landau-Lifshitz
(LL) equation, augmented by the spin-current term:
\begin{equation}
\hbar\frac{\partial\mathbf{S}_{i}}{\partial t}+\hbar\left(\mathbf{v}_{S}\cdot\nabla\right)\mathbf{S}_{i}=\mathbf{S}_{i}\times\mathbf{H}_{\mathrm{eff},S,i},-\alpha\mathbf{S}_{i}\times\left(\mathbf{S}_{i}\times\mathbf{H}_{\mathrm{eff},S,i}\right)\label{LL}
\end{equation}
where $\mathbf{H}_{\mathrm{eff},S,i}=-\partial\mathcal{H}/\partial\mathbf{S}_{i}$
is the effective field (in the energy units) acting on the TM sublattice,
$\alpha$ is the phenomenological damping coefficient, and $\mathbf{v}_{S}$
is the convective velocity arising from the (time-dependent) electric
current flowing in the film's plane and acting on the TM spins $\mathbf{S}_{i}$.
The electric current is initially non-polarized and it flows through
the material carrying the spin polarization from one site to the other.
The equation of motion for the RE spins is similar, except $\mathbf{v}_{\Sigma}$
is negligible because RE spins are due to the $f$-electrons which
are close to the atomic cores and effectively decoupled from the charge
carriers. For the gradient of $\mathbf{S}_{i}$ we use the finite-difference
form on the lattice. This model differs from that including the spin-polarized
current injected in the system.

\begin{figure}
\begin{centering}
\includegraphics[width=8cm]{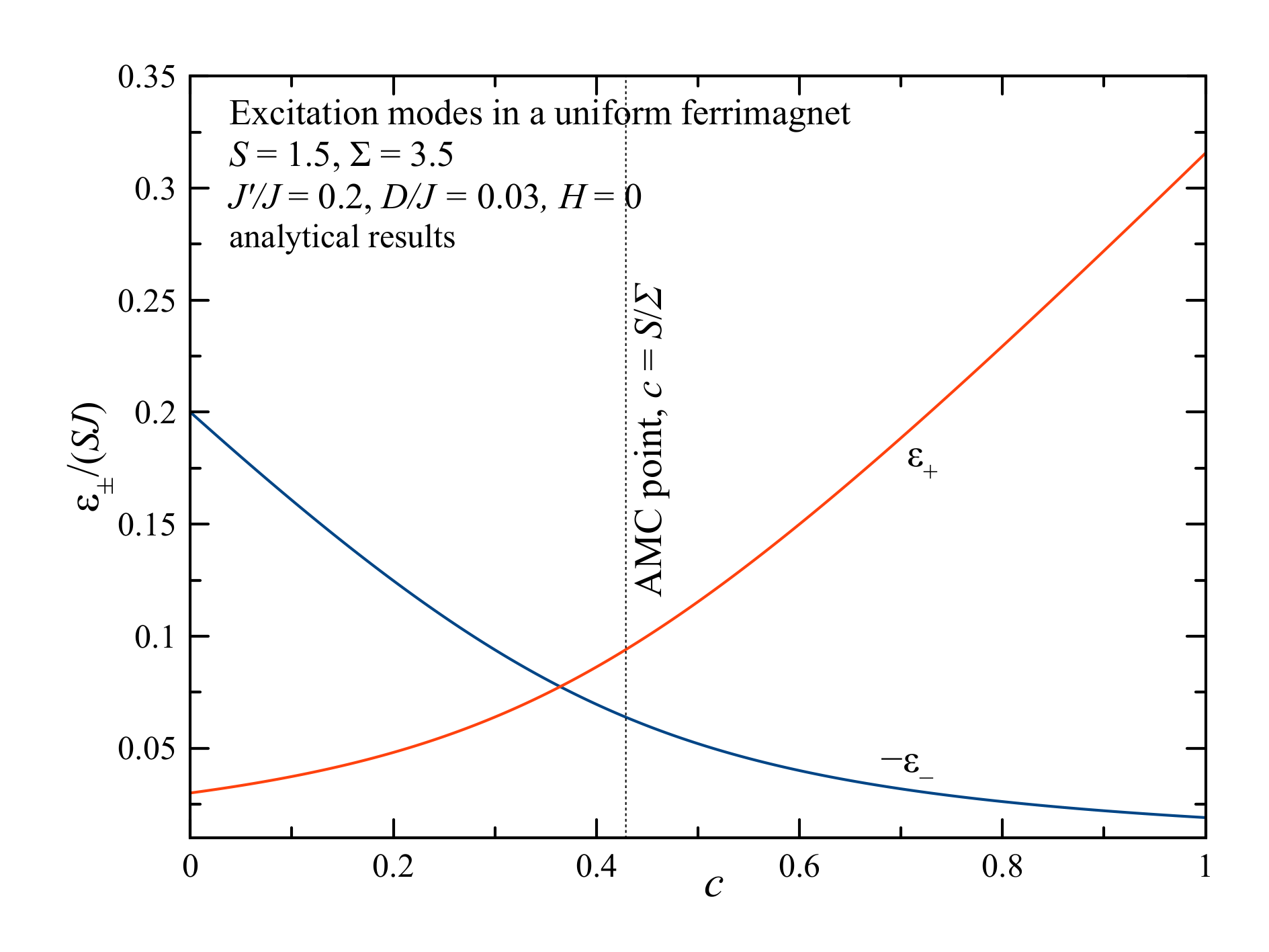}
\par\end{centering}
\caption{Energies of the uniform modes in a ferrimagnet vs the RE concentration
$c$, as given by Eq. (\ref{epsilon_pm-general}).}\label{Fig_eps_pm_vs_c_analytical}
\end{figure}

In the uniform state, the ferrimagnet possesses two spin-wave branches.
Replacing $p_{i}\Rightarrow c$ (the no-disorder model) and using
the linear spin-wave theory, for the energies of the modes $\varepsilon_{\pm}=\omega_{\pm}/\hbar$
one obtains \citep{DG-EC-PRB2026}
\begin{equation}
\varepsilon_{\pm}=\frac{1}{2}\left[\varepsilon_{TM}+\varepsilon_{RE}\pm\sqrt{\left(\varepsilon_{TM}-\varepsilon_{RE}\right)^{2}-4c\Sigma SJ'{}^{2}}\right],\label{epsilon_pm-general}
\end{equation}
where
\begin{eqnarray}
\varepsilon_{TM} & \equiv & S\left(J_{0}-J_{\mathbf{k}}+D\right)+g\mu_{B}H+c\Sigma J'\\
\varepsilon_{RE} & \equiv & g'\mu_{B}H-J'S,
\end{eqnarray}
and $J_{\mathbf{k}}$ is the lattice Fourier transform of $J_{ij}$
and $J_{0}\equiv J_{\mathbf{k=0}}$. The signs of $\varepsilon_{\pm}$
are unimportant as they depend on the direction of spin precession.
The DMI does not affect the linear excitation modes in the uniform
state. Figure \ref{Fig_eps_pm_vs_c_analytical} shows the RE-concentration
dependence of both modes at $\mathbf{k=0}$. 

For the $H=0$, the uniform modes cross at 
\begin{equation}
c=\frac{S}{\Sigma}\left(1-\frac{D}{J'}\right),
\end{equation}
where
\begin{equation}
\varepsilon_{\pm}=\pm S\sqrt{DJ'}.
\end{equation}
At the angular-momentum compensation (AMC) point, $c=S/\Sigma$, one
has
\begin{equation}
\varepsilon_{\pm}=\frac{S}{2}\left(D\pm\sqrt{D\left(4J'+D\right)}\right).
\end{equation}
Typically $J'\gg D$, so that the second term under the square root
is negligible. One can see that the mode splitting at compensation
is $\varepsilon_{+}-\left|\varepsilon_{-}\right|=SD.$ For the chosen
parameters one has $\varepsilon_{+}/(SJ)=0.0925$ and $\varepsilon_{-}/(SJ)=-0.0625$. 

Far from the AMC point, one can expand Eq. (\ref{epsilon_pm-general})
considering $J'$ as large compared to $D$. One obtains
\begin{equation}
\varepsilon_{\mathrm{LF}}\cong\frac{S^{2}\left(J_{0}-J_{\mathbf{k}}+D\right)+\left(gS-cg'\Sigma\right)\mu_{B}H}{S-c\Sigma}
\end{equation}
for the low-frequency (acoustical) mode and $\varepsilon_{\mathrm{HF}}\cong\left(c\Sigma-S\right)J'$
for the high-frequency (exchange or optical) mode. Note that $\varepsilon_{\mathrm{HF}}=\varepsilon_{-}$
and $\varepsilon_{\mathrm{LF}}=\varepsilon_{+}$ for smaller $c$
and $\varepsilon_{\mathrm{HF}}=\varepsilon_{+}$ and $\varepsilon_{\mathrm{LF}}=\varepsilon_{-}$
for larger $c$, see Fig. \ref{Fig_eps_pm_vs_c_analytical}.

\section{Thiele equations for rigid TM and RE skyrmions}

\label{Sec_Thiele-equations}

\begin{figure}
\begin{centering}
\includegraphics[width=8cm]{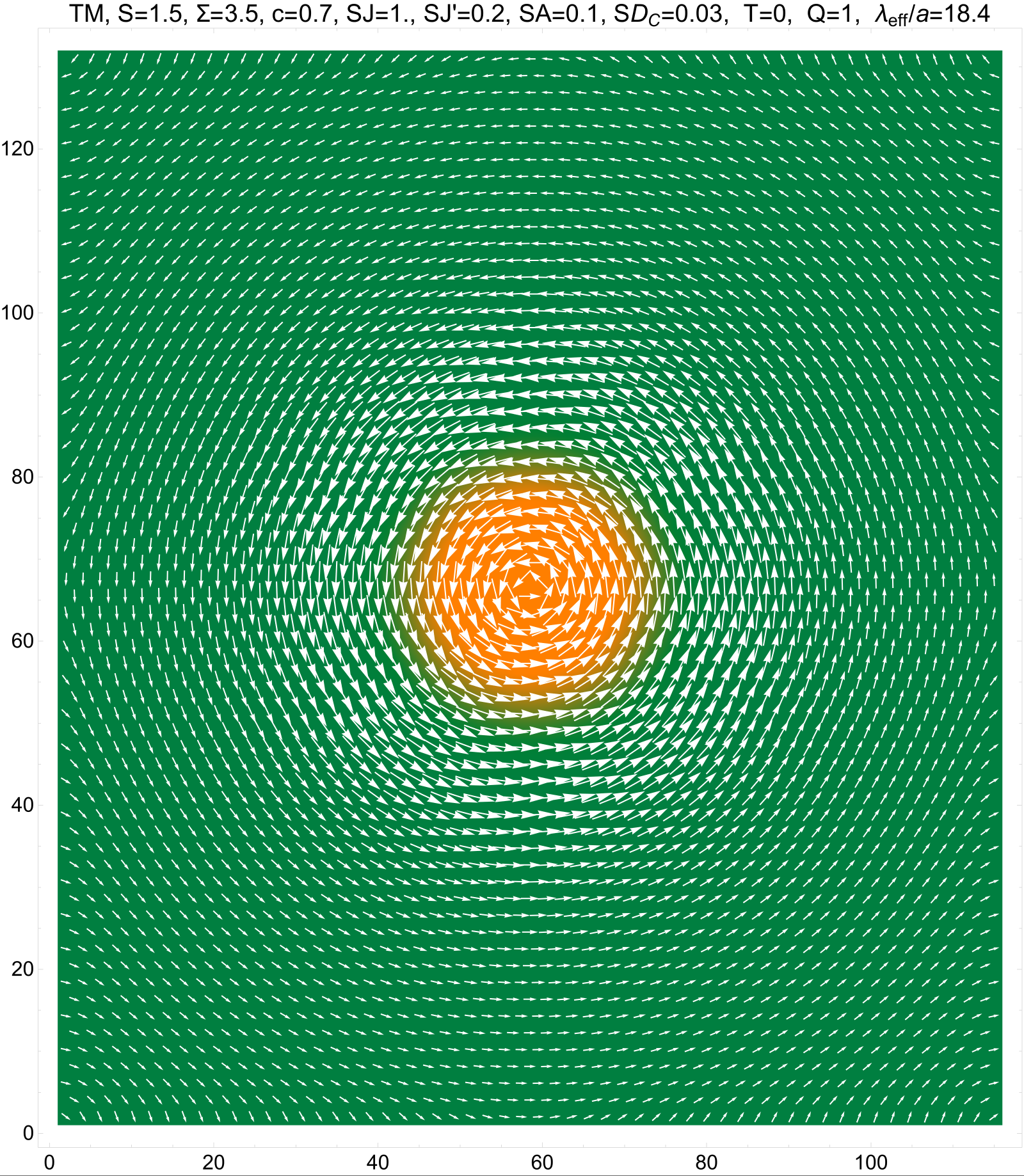}
\par\end{centering}
\begin{centering}
\includegraphics[width=8cm]{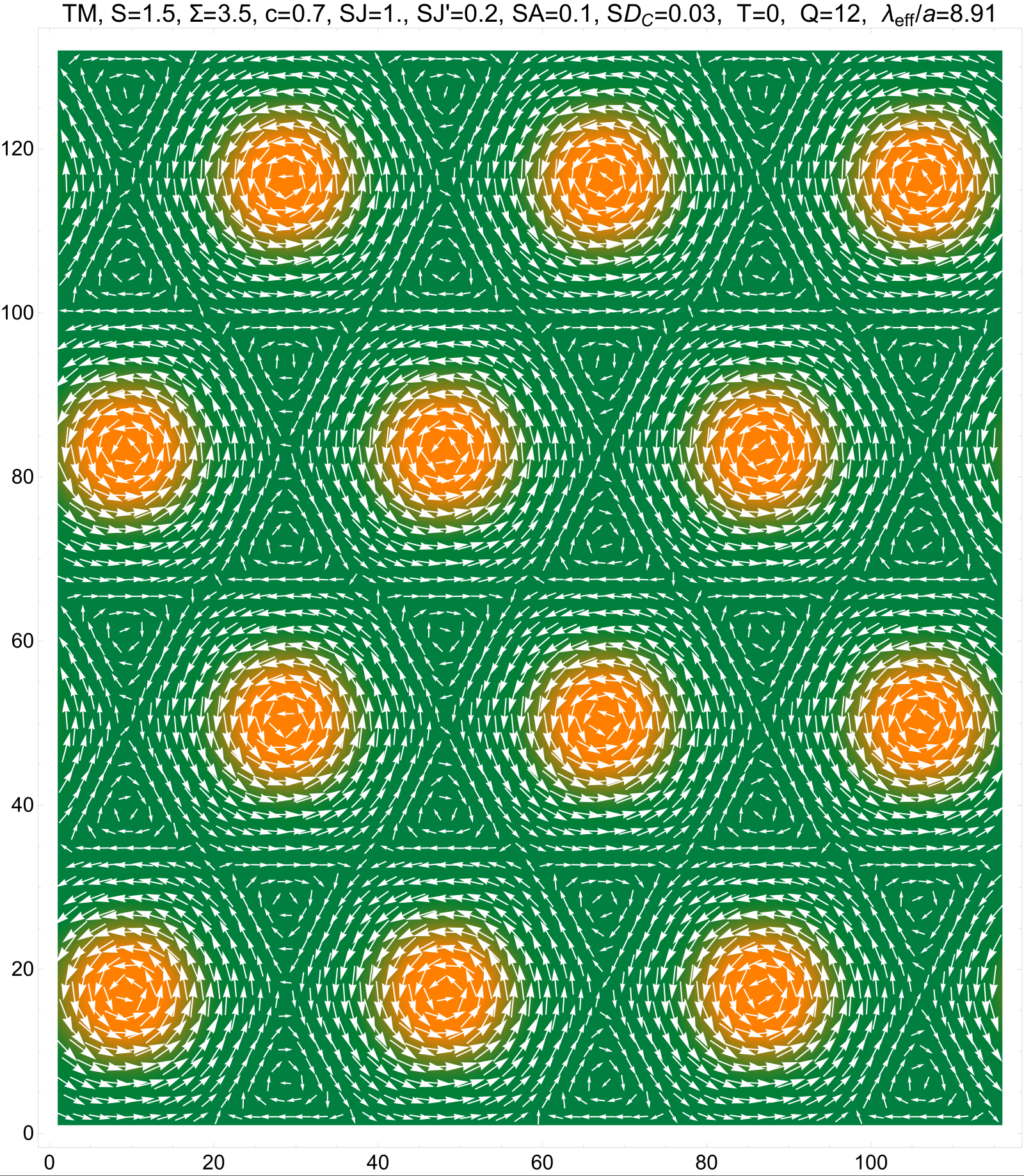}
\par\end{centering}
\caption{The $116\times132$ of spins with a SS (top) and a SkL (bottom), with
only TM spins shown. Orange/green -- spins up/down. White arrows
are in-plane spin components. }\label{Fig_SkL}
\end{figure}

Whereas the dynamics of skyrmions can be computed at the atomic level
using Eq. (\ref{LL}), analytically it can be described by the Thiele
equation \citep{Thiele} that has a topological background and provides
a linear relation between the skyrmion velocity and the applied force,
thus no inertia effects. There are at least three derivations of the
Thiele equation. Two of them assume a rigid shape of the skyrmions
and use (i) the Lagrangian approach and (ii) the LL equation in the
Gilbert form (see, e.g., Ref. \citep{capgarchu2025}). It was argued
that the deformation of the skyrmion gives rise to inertia effects,
and a phenomenological skyrmion-mass term was added to the Thiele
equation. However, microscopic calculations, both analytical and numerical,
show that these inertia effects are very small, at least for small
skyrmions \citep{garchu-unpub}. On the other hand, the third derivation
of the Thiele equation using the definition of the skyrmion's center
$\mathbf{R}$ as the center of mass of its topological charge \citep{Komineas2015,EC-DA-EPL2025}
\begin{equation}
{\bf R}=\frac{1}{Q}\int dxdy\,\mathbf{r}q(x,y),\label{Skyrmion_center_R_def}
\end{equation}
where 
\begin{equation}
q=\frac{1}{4\pi}\left(\frac{\partial{\bf s}}{\partial x}\times\frac{\partial{\bf s}}{\partial y}\right)\cdot{\bf s}\label{Topological_charge_q_def}
\end{equation}
is the topological charge density and $Q=\int dxdy\,q(x,y)=0,\pm1,\ldots$
is the topological charge of the skyrmion, is exact and insensitive
to any skyrmion's deformations. In our case of skyrmions created by
the DMI, one has $|Q|=1$. As the difference between the positions
of the skyrmion's center defined in different ways is much smaller
than the skyrmion's size $\lambda$, it is clear that the skyrmion's
trajectory cannot be significantly changed by using another definition
of the skyrmion's center, sensitive to the skyrmion deformations.
Thus, inertia effects for a ferromagnetic skyrmion are small or absent,
and here is no place for the mass term in the Thiele equation. This
is a consequence of the skyrmion having the topological charge. Conversely,
the domain wall in a biaxial ferromagnet does not have a topological
charge but has a mass.

For a ferrimagnet, there are skyrmions in the TM and RE subsystems,
and the shapes of them are assumed to be given by \citep{Panigrahy2022,Nowak-PRB2023,Lau2025,EC-DG-arXiv}
\begin{equation}
\mathbf{S}(\mathbf{r})=S\mathbf{f}\left(\mathbf{r}-\mathbf{R}_{S}\right),\qquad\boldsymbol{\sigma}(\mathbf{r})=-\Sigma\mathbf{f}\left(\mathbf{r}-\mathbf{R}_{\Sigma}\right),\label{Skyrmions_shapes}
\end{equation}
where $\mathbf{f}$ is a unit-vector function directed up in the skyrmion's
core and down at infinity. We anticipate that dynamically the skyrmions'
centers $\mathbf{R}_{S,\Sigma}$ can be displaced by the vector $\mathbf{d}$
with respect to each other:
\begin{equation}
\mathbf{d\equiv}\mathbf{R}_{S}-\mathbf{R}_{\Sigma},\label{d_def}
\end{equation}
and we neglect deformations of the TM and RE skyrmions \citep{Panigrahy2022,Nowak-PRB2023,Lau2025,EC-DG-arXiv}.
This approximation is justified by the accuracy of the Thiele equation
and by the fact that the interaction force between the two skyrmions
is an integral quantity in which the effect of deformations should
be averaged out to the lowest order. 

The system of two coupled Thiele equations \citep{Thiele} for the
skyrmions in the TM and RE subsystems, augmented by the spin-current
terms, has the form
\begin{eqnarray}
\mathbf{\bar{F}}_{S} & = & -\mathbf{G}_{S}\times\left(\mathbf{V}_{S}-\mathbf{v}_{S}\right)+\Gamma_{S}\left(\mathbf{V}_{S}-\mathbf{v}_{S}\right)\nonumber \\
\mathbf{\bar{F}}_{\Sigma} & = & \mathbf{G}_{\Sigma}\times\left(\mathbf{V}_{\Sigma}-\mathbf{v}_{\Sigma}\right)+\Gamma_{\Sigma}\left(\mathbf{V}_{\Sigma}-\mathbf{v}_{\Sigma}\right).\label{Thiele}
\end{eqnarray}
Here $\mathbf{\bar{F}}_{S}$ and $\mathbf{\bar{F}}_{\Sigma}$ are
total forces acting on the TM and RE skyrmions, $\mathbf{G}_{S}=G_{S}\mathbf{e}_{z}$
and $\mathbf{G}_{\Sigma}=G_{\Sigma}\mathbf{e}_{z}$ are gyrovectors,
\begin{equation}
G_{S,\Sigma}=4\pi\hbar\rho_{S,\Sigma},\qquad\rho_{S}=S/a^{2},\quad\rho_{\Sigma}=c\Sigma/a^{2},\label{G_S_def}
\end{equation}
$\rho_{S,\Sigma}$ are 2D spin densities defined in Eq. (\ref{Energy_density}).
$\Gamma_{S,\Sigma}=\alpha\mathcal{D}\hbar\rho_{S,\Sigma}$ are damping
coefficients (see, e.g., the Appendix in Ref. \citep{capgarchu2025})
where
\begin{equation}
\mathcal{D}\equiv\frac{1}{2}\int dxdy\left[\left(\frac{\partial\mathbf{f}}{\partial x}\right)^{2}+\left(\frac{\partial\mathbf{f}}{\partial y}\right)^{2}\right].\label{Dcal_def}
\end{equation}
For the BP skyrmion, which is realized for a small DMI and small anisotropy,
the exchange dominates and $\mathcal{D}$ is equal to $4\pi|Q|$,
so that 
\begin{equation}
\frac{\Gamma_{S}}{G_{S}}=\frac{\Gamma_{\Sigma}}{G_{\Sigma}}=\frac{\alpha\mathcal{D}}{4\pi}\equiv\Lambda\label{Lambda_def}
\end{equation}
is of the order of the damping constant $\alpha$ (for simplicity,
we assume the same $\alpha$ in the TM and RE subsystems). Note that
operating with the RE spin density $c\Sigma/a^{2}$ we neglect the
effect of disorder. Although $\mathbf{v}_{\Sigma}$ is negligible,
we keep it for generality.

The forces acting on the skyrmions are given by 
\begin{eqnarray}
\mathbf{\bar{F}}_{S} & = & \mathbf{F}_{S}+\mathbf{F}_{S\Sigma}\nonumber \\
\mathbf{\bar{F}}_{\Sigma} & = & \mathbf{F}_{\Sigma}+\mathbf{F}_{\Sigma S},\label{Forces}
\end{eqnarray}
where $\mathbf{F}_{S}$ and $\mathbf{F}_{\Sigma}$ are external forces,
whereas $\mathbf{F}_{S\Sigma}$ and $\mathbf{F}_{\Sigma S}$ are the
interaction forces. The former can be, e.g., due to the interaction
of the skyrmion with boundaries or due to the magnetic-field gradient.
The latter follow from the TM-RE skyrmion interaction energy $U_{S\Sigma}$.
Replacing summation by integration for $\lambda\gg a$, one obtains
\begin{eqnarray}
U_{S\Sigma} & = & \frac{J'}{a^{2}}\int dxdy\mathbf{S}(\mathbf{r})\cdot\boldsymbol{\Sigma}(\mathbf{r})\nonumber \\
 & = & -\rho_{S}\rho_{\Sigma}\mathcal{J}'\int dxdy\mathbf{f}(\mathbf{r})\cdot\mathbf{f}\left(\mathbf{r}+\mathbf{d}\right),\label{USSigma_def}
\end{eqnarray}
where $\mathcal{J}'\equiv a^{2}J'$ is the intersublattice coupling
of spin densities in Eq. (\ref{Energy_density}). Expanding this for
$d\ll\lambda$, one obtains
\begin{equation}
U_{S\Sigma}=\frac{1}{2}\rho_{S}\rho_{\Sigma}\mathcal{J}'\mathcal{D}d^{2}+\mathrm{const},\label{USSigma_res}
\end{equation}
where $\mathcal{D}$ is given by Eq. (\ref{Dcal_def}). Notice that
the exchange energy associated with the separation of the centers
of sublattice skyrmions is of the order of $J'(d/a)^{2}$, so that
in practice one should always expect $d\ll a$. Now the interaction
forces are given by 
\begin{equation}
\mathbf{F}_{S\Sigma}=-\mathbf{F}_{\Sigma S}=-\frac{\partial U_{S\Sigma}}{\partial\mathbf{R}_{S}}=-\rho_{S}\rho_{\Sigma}\mathcal{J}'\mathcal{D}\mathbf{d}.\label{FSSigma_res}
\end{equation}

\section{Skyrmion mass and cyclotron resonance}

\label{Sec_Skyrmion-mass}

\begin{figure}
\begin{centering}
\includegraphics[width=8cm]{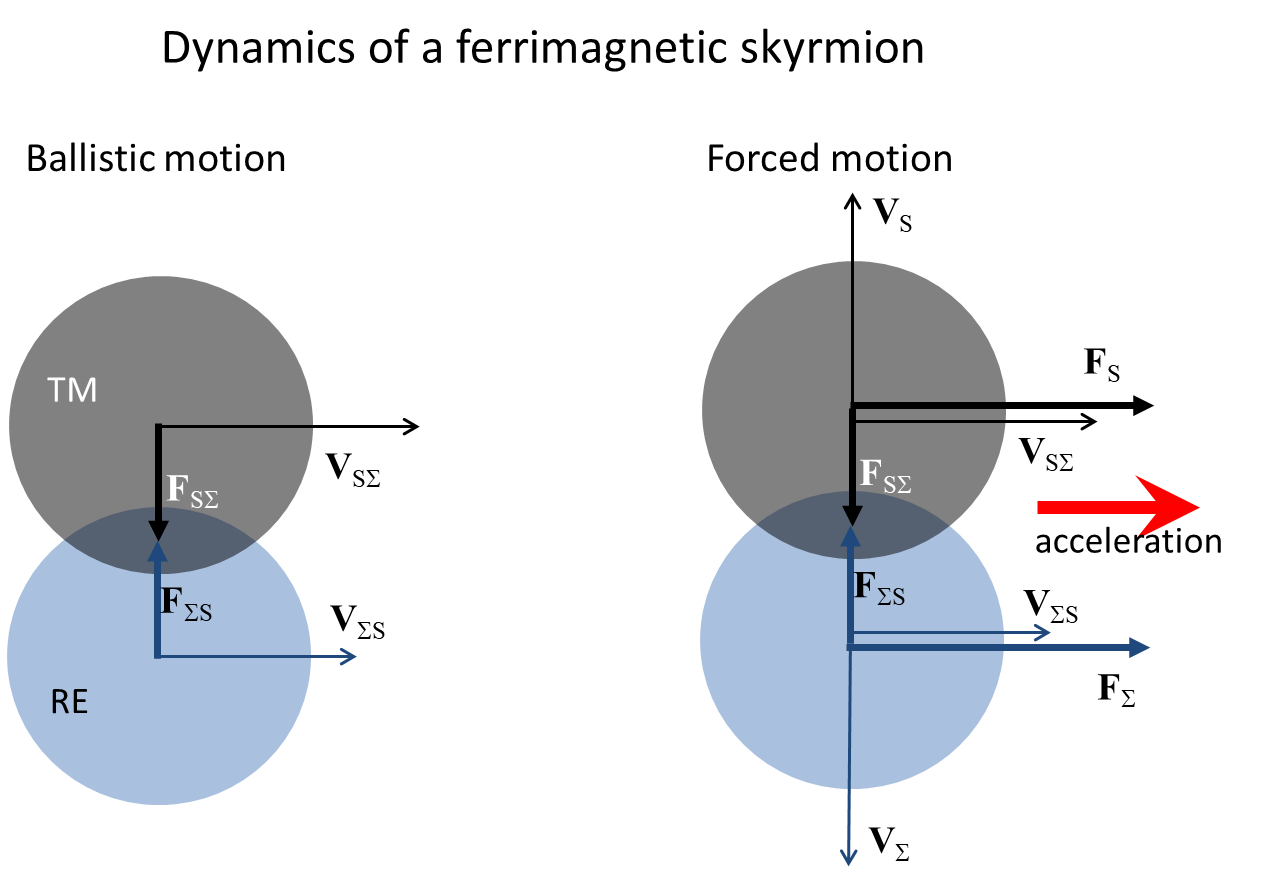}
\par\end{centering}
\caption{Dynamics of a ferrimagnetic skyrmion explained. The forces $\mathbf{F}_{S,\Sigma,\ldots}$
are related to the velocities $\mathbf{V}_{S,\Sigma,\ldots}$ by the
Thiele equation. On the left, shifting of the TM and RE skyrmions
(exaggerated) gives rise to the pair of interaction forces $\mathbf{F}_{S\Sigma}$
and $\mathbf{F}_{\Sigma S}$ that result in the ballistic motion of
both skyrmions in the same perpendicular direction. Since in general
the velocities $\mathbf{V}_{S\Sigma}$ and $\mathbf{V}_{\Sigma S}$
are different, the TM and RE skyrmions are circling each other. On
the right, external forces $\mathbf{F}_{S}$ and $\mathbf{F}_{\Sigma}$
cause the increasing skyrmion shift that, in turn, results in the
accelerated motion in the direction of the applied forces. }\label{Fig_Dynamics}

\end{figure}

The interaction between the shifted TM and RE skyrmions results in
their motion in the same direction, that is, in the ballistic motion
of the system of the TM-RE skyrmions. Away from the angular-momentum
compensation point, the velocities of the TM and RE skyrmions are
different, thus the motion of the ferrimagnetic skyrmion is curved.
If an external force is applied, the skyrmion shift $d$ increases
with time and the system accelerates in the direction of the applied
force. The dynamics of the ferrimagnetic skyrmion is explained in
Fig. \ref{Fig_Dynamics}.

Below, we will obtain a massive equation of motion for a ferrimagnetic
skyrmion by rearranging the two Thiele equations into a single equation
for a combination of the skyrmions' velocities $\mathbf{V}_{S,\Sigma}=\mathbf{\dot{R}}_{S,\Sigma}$.
Resolving Eq. (\ref{Thiele}) for the velocities, one obtains
\begin{eqnarray}
\mathbf{V}_{S}-\mathbf{v}_{S} & = & \frac{\mathbf{G}_{S}\times\mathbf{\bar{F}}_{S}+\Gamma_{S}\mathbf{\bar{F}}_{S}}{G_{S}^{2}+\Gamma_{S}^{2}}\nonumber \\
\mathbf{V}_{\Sigma}-\mathbf{v}_{\Sigma} & = & \frac{-\mathbf{G}_{\Sigma}\times\mathbf{\bar{F}}_{\Sigma}+\Gamma_{\Sigma}\mathbf{\bar{F}}_{\Sigma}}{G_{\Sigma}^{2}+\Gamma_{\Sigma}^{2}}.
\end{eqnarray}
Separating the terms with the external forces and spin currents and
using Eq. (\ref{FSSigma_res}), one obtains
\begin{eqnarray}
\mathbf{V}_{S}-\mathbf{U}_{S} & = & -\rho_{S}\rho_{\Sigma}\mathcal{J}'\mathcal{D}\frac{\mathbf{G}_{S}\times\mathbf{d}+\Gamma_{S}\mathbf{d}}{G_{S}^{2}+\Gamma_{S}^{2}}\\
\mathbf{V}_{\Sigma}-\mathbf{U}_{\Sigma} & = & -\rho_{S}\rho_{\Sigma}\mathcal{J}'\mathcal{D}\frac{\mathbf{G}_{\Sigma}\times\mathbf{d}-\Gamma_{\Sigma}\mathbf{d}}{G_{\Sigma}^{2}+\Gamma_{\Sigma}^{2}},
\end{eqnarray}
where
\begin{eqnarray}
\mathbf{U}_{S} & \equiv & \mathbf{v}_{S}+\frac{\mathbf{G}_{S}\times\mathbf{F}_{S}+\Gamma_{S}\mathbf{F}_{S}}{G_{S}^{2}+\Gamma_{S}^{2}}\nonumber \\
\mathbf{U}_{\Sigma} & \equiv & \mathbf{v}_{\Sigma}+\frac{-\mathbf{G}_{\Sigma}\times\mathbf{F}_{\Sigma}+\Gamma_{\Sigma}\mathbf{F}_{\Sigma}}{G_{\Sigma}^{2}+\Gamma_{\Sigma}^{2}}\label{US_USig_def-1}
\end{eqnarray}
are external terms. Obtaining the equation of motion for the skyrmions'
velocities is possible within the vector formalism in the absence
of damping. In the presence of damping, one needs to use the more
powerful matrix formalism and to rewrite the resolved Thiele equations
above as
\begin{equation}
\mathbf{V}_{S,\Sigma}-\mathbf{U}_{S,\Sigma}=-\rho_{S}\rho_{\Sigma}\mathcal{J}'\mathcal{D}\mathbb{M}_{S,\Sigma}\cdot\mathbf{d}
\end{equation}
where
\begin{eqnarray}
\mathbb{M}_{S} & = & \frac{1}{G_{S}^{2}+\Gamma_{S}^{2}}\left(\begin{array}{cc}
\Gamma_{S} & -G_{S}\\
G_{S} & \Gamma_{S}
\end{array}\right)=\left(\begin{array}{cc}
\Gamma_{S} & G_{S}\\
-G_{S} & \Gamma_{S}
\end{array}\right)^{-1}\nonumber \\
\mathbb{M}_{\Sigma} & = & \frac{1}{G_{\Sigma}^{2}+\Gamma_{\Sigma}^{2}}\left(\begin{array}{cc}
-\Gamma_{\Sigma} & -G_{\Sigma}\\
G_{\Sigma} & -\Gamma_{\Sigma}
\end{array}\right)\label{M_matr_def}
\end{eqnarray}
or, with the use of Eq. (\ref{Lambda_def}),
\begin{eqnarray}
\mathbb{M}_{S} & = & \frac{1}{G_{S}\left(1+\Lambda^{2}\right)}\left(\begin{array}{cc}
\Lambda & -1\\
1 & \Lambda
\end{array}\right)\nonumber \\
\mathbb{M}_{\Sigma} & = & \frac{1}{G_{\Sigma}\left(1+\Lambda^{2}\right)}\left(\begin{array}{cc}
-\Lambda & -1\\
1 & -\Lambda
\end{array}\right).\label{M_matr_Lam}
\end{eqnarray}
Also, 
\begin{equation}
\mathbf{U}_{S}\equiv\mathbf{v}_{S}+\mathbb{M}_{S}\cdot\mathbf{F}_{S},\qquad\mathbf{U}_{\Sigma}\equiv\mathbf{v}_{\Sigma}-\mathbb{M}_{\Sigma}\cdot\mathbf{F}_{\Sigma}.\label{US_USig_def}
\end{equation}

\begin{figure}
\begin{centering}
\includegraphics[width=8cm]{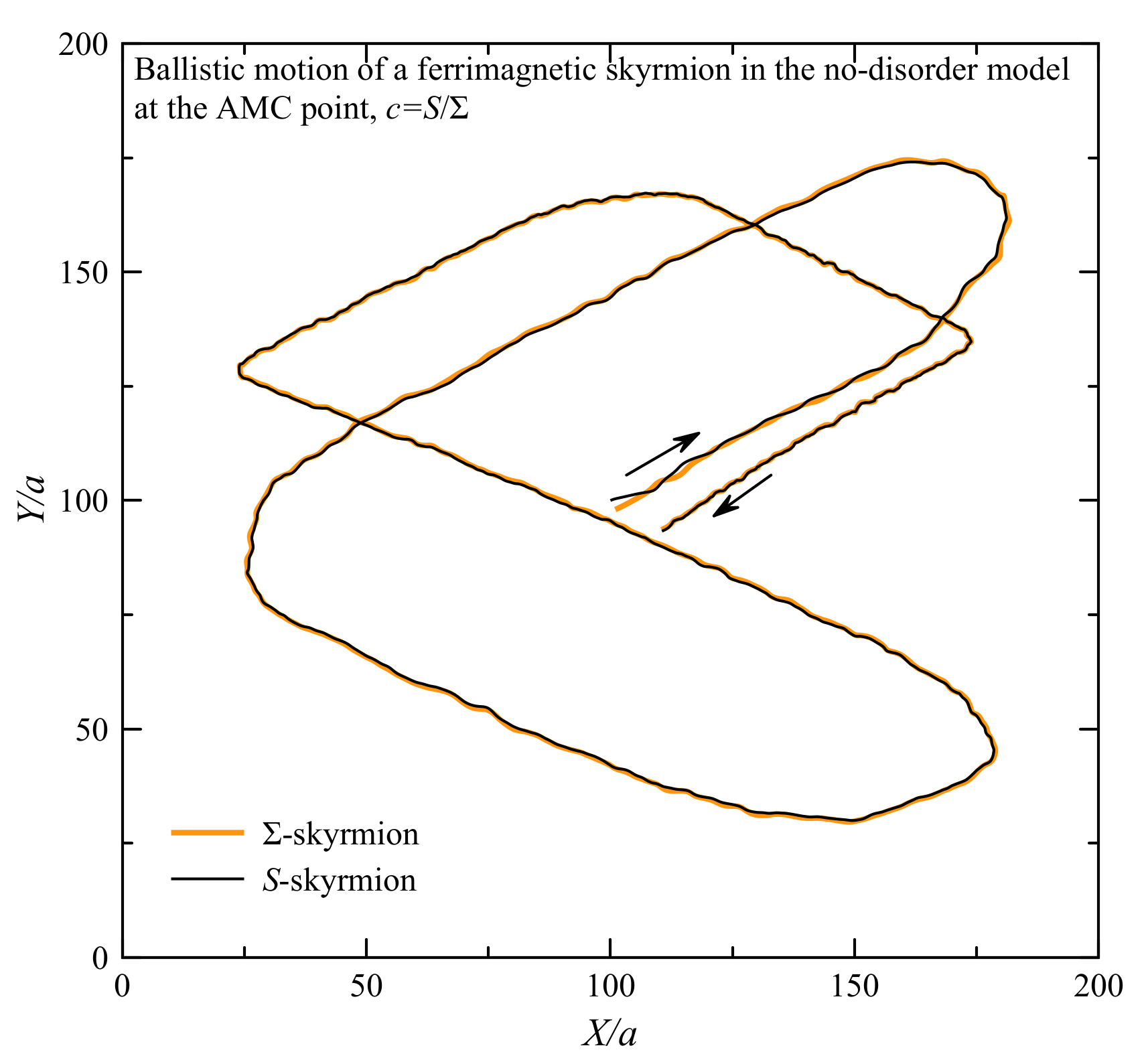}
\par\end{centering}
\caption{Ballistic motion of the ferrimagnetic skyrmion in the no-disorder
model at the AMC point, showing reflections from the boundaries. In
the initial state, RE and TM skyrmions are shifted with respect to
each other to initiate their motion. }\label{Fig_Ballistic_motion}
\end{figure}

\begin{figure}
\begin{centering}
\includegraphics[width=8cm]{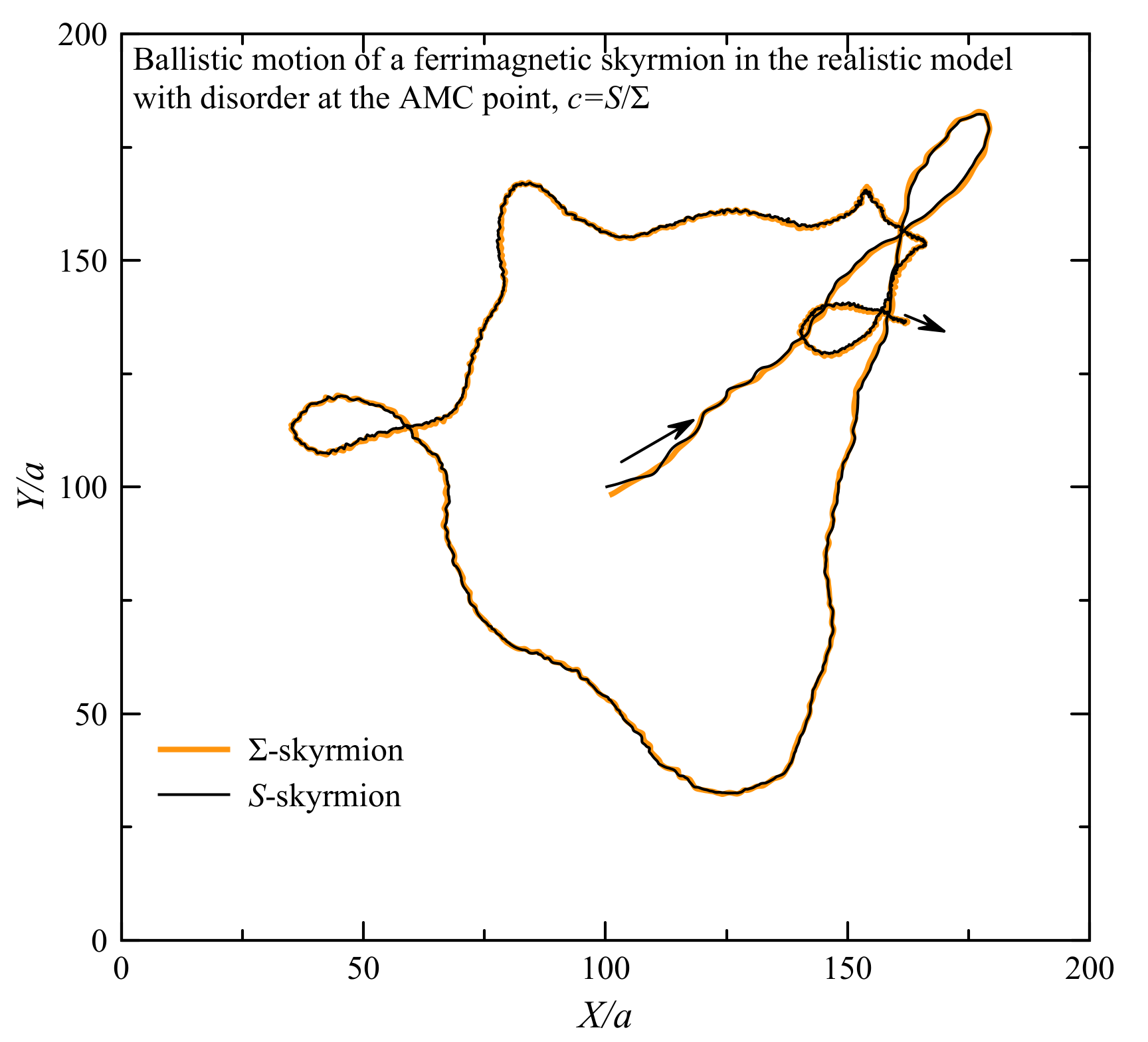}
\par\end{centering}
\begin{centering}
\includegraphics[width=8cm]{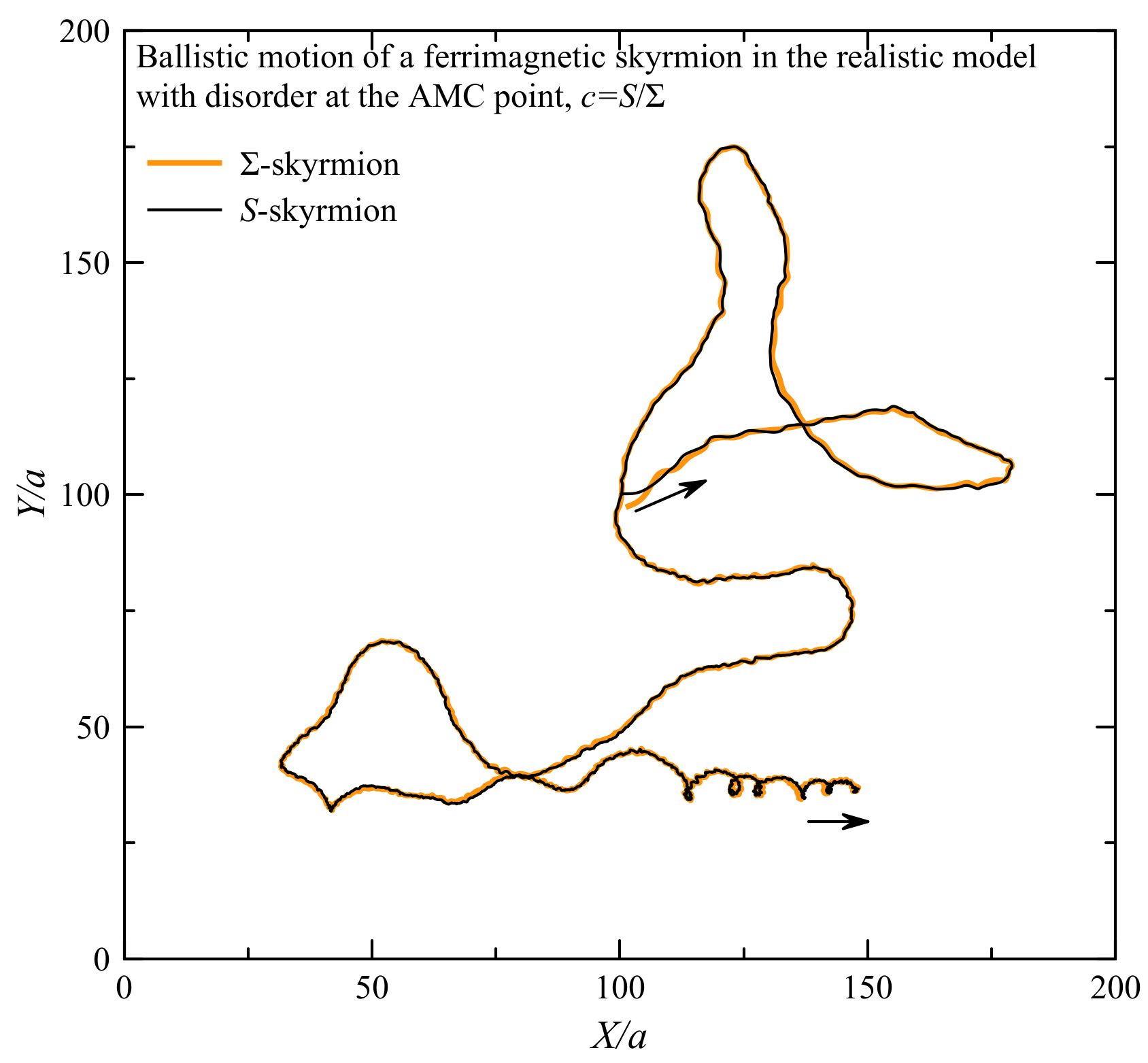}
\par\end{centering}
\caption{Ballistic motion of the ferrimagnetic skyrmion in the realistic model
with disorder at the AMC point. Top and bottom: different realizations
of the disorder. }\label{Fig_Ballistic_motion_disorder}
\end{figure}

Next, we define the weighted position of the skyrmion's center:
\begin{equation}
\mathbf{R}=p_{S}\mathbf{R}_{S}+p_{\Sigma}\mathbf{R}_{\Sigma},\label{R_weighted_def}
\end{equation}
where $p_{S}+p_{\Sigma}=1$. Similarly, one defines $\mathbf{V}=\mathbf{\dot{R}}$
and the time derivative of $\mathbf{V}$. For $\mathbf{V}$ one has
\begin{equation}
\mathbf{V}=\mathbf{U}-\rho_{S}\rho_{\Sigma}\mathcal{J}'\mathcal{D}\left(p_{S}\mathbb{M}_{S}+p_{\Sigma}\mathbb{M}_{\Sigma}\right)\cdot\mathbf{d},\label{V_weighted}
\end{equation}
where, similarly, $\mathbf{U}=p_{S}\mathbf{U}_{S}+p_{\Sigma}\mathbf{U}_{\Sigma}$.
On the other hand, 
\begin{equation}
\mathbf{\dot{d}}=\mathbf{V}_{S}-\mathbf{V}_{\Sigma}=\mathbf{U}_{S}-\mathbf{U}_{\Sigma}-\rho_{S}\rho_{\Sigma}\mathcal{J}'\mathcal{D}\left(\mathbb{M}_{S}-\mathbb{M}_{\Sigma}\right)\cdot\mathbf{d}.\label{ddot_eq}
\end{equation}
Manipulating these two equations, one can obtain a closed equation
of motion for $\mathbf{V}$ as follows. One differentiates Eq. (\ref{V_weighted})
over time and substitutes $\mathbf{\dot{d}}$ given by Eq. (\ref{ddot_eq}),
thus relating $\mathbf{\dot{V}}$ to $\mathbf{d}$. In the resulting
equation, one substitutes $\mathbf{d}$ expressed via $\mathbf{V}$
by the resolution of Eq. (\ref{V_weighted}). The resulting equation
of motion has the form
\begin{equation}
M\dot{\mathbf{V}}+\mathbb{G}\cdot\mathbf{V}=\mathbf{K},\label{Newt_eq_matrix}
\end{equation}
where the mass is given by
\begin{equation}
M=\frac{G_{S}G_{\Sigma}\left(1+\Lambda^{2}\right)}{\rho_{S}\rho_{\Sigma}\mathcal{J}'\mathcal{D}}=\frac{\left(4\pi\hbar\right)^{2}\left(1+\Lambda^{2}\right)}{\mathcal{J}'\mathcal{D}},\label{M_damped}
\end{equation}
the matrix $\mathbb{G}$ has the form
\begin{eqnarray}
\mathbb{G} & = & \left(1+\Lambda^{2}\right)G_{S}G_{\Sigma}\left(p_{S}\mathbb{M}_{S}+p_{\Sigma}\mathbb{M}_{\Sigma}\right)\nonumber \\
 &  & \cdot\left(\mathbb{M}_{S}-\mathbb{M}_{\Sigma}\right)\cdot\left(p_{S}\mathbb{M}_{S}+p_{\Sigma}\mathbb{M}_{\Sigma}\right)^{-1}.
\end{eqnarray}
This simplifies to
\begin{equation}
\mathbb{G}=\left(\begin{array}{cc}
\Gamma_{S}+\Gamma_{\Sigma} & G_{S}-G_{\Sigma}\\
-\left(G_{S}-G_{\Sigma}\right) & \Gamma_{S}+\Gamma_{\Sigma}
\end{array}\right),\label{G_matr_result}
\end{equation}
which is independent of the weights $p_{S,\Sigma}$. The external
term $\mathbf{K}$ after matrix multiplications becomes
\begin{equation}
\mathbf{K}=M\dot{\mathbf{U}}+\mathbf{F}_{S}+\mathbf{F}_{\Sigma}+\Gamma_{S}\mathbf{v}_{S}+\Gamma_{\Sigma}\mathbf{v}_{\Sigma}-\mathbf{G}_{S}\times\mathbf{v}_{S}+\mathbf{G}_{\Sigma}\times\mathbf{v}_{\Sigma}.\label{K_def}
\end{equation}
Finally, Eq. (\ref{Newt_eq_matrix}) can be rewritten in the vector
form as
\begin{equation}
M\dot{\mathbf{V}}-\left(\mathbf{G}_{S}-\mathbf{G}_{\Sigma}\right)\times\mathbf{V}+\left(\Gamma_{S}+\Gamma_{\Sigma}\right)\mathbf{V}=\mathbf{K}.\label{Newton_eq_damped}
\end{equation}
 The expression for the skyrmion mass $M$ was obtained from the condition
that $\mathbf{K}$ contains the total external force $\mathbf{F}_{S}+\mathbf{F}_{\Sigma}$
without any coefficient. The gyration term in this equation defines
to the cyclotron frequency vector
\begin{equation}
\boldsymbol{\Omega}=\frac{\mathbf{G}_{S}-\mathbf{G}_{\Sigma}}{M}=\frac{J'}{\hbar}\frac{\mathcal{D}}{4\pi}\left|S-c\Sigma\right|\mathbf{e}_{z}.\label{Omega_cyclotron}
\end{equation}
 The cyclotron frequency simplifies to 
\begin{equation}
\Omega=\frac{J'}{\hbar}\left|S-c\Sigma\right|\label{Omega_cyclotron-BP}
\end{equation}
in the BP limit, $\mathcal{D}=4\pi$. The SCR frequency vanishes at
the AMC point. Correspondingly, the cyclotron radius $R_{\mathrm{cyc}}=V/\Omega$
diverges. In this case, for the no-disorder model, the skyrmion can
perform a straight ballistic motion with reflections from the boundaries
of the system. This motion, obtained by the numerical solution in
terms of lattice spins with the rigid-wall boundary condition is illustrated
in Fig. \ref{Fig_Ballistic_motion}. However, for the realistic model
with disorder, ballistic motion of the skyrmion follows a random trajectory
depending on the distribution of the RE atoms in the system, see Fig.
\ref{Fig_Ballistic_motion_disorder}. It is remarkable that for $\mathcal{D}=4\pi$
the only model parameter in the cyclotron frequency is the intersublattice
coupling $J'$. The initial condition to Eq. (\ref{Newton_eq_damped})
is given by Eq. (\ref{V_weighted}), that is, by the Thiele equation.
Using Eq. (\ref{M_damped}), it can be rewritten in the form
\begin{equation}
\mathbf{V}=\mathbf{U}-\frac{G_{S}G_{\Sigma}\left(1+\Lambda^{2}\right)}{M}\left(p_{S}\mathbb{M}_{S}+p_{\Sigma}\mathbb{M}_{\Sigma}\right)\cdot\mathbf{d}\label{Init_cond_general}
\end{equation}
and further, using Eq. (\ref{M_matr_Lam}), as
\begin{equation}
\mathbf{V}=\mathbf{U}+\frac{1}{M}\left(\begin{array}{cc}
A & B\\
-B & A
\end{array}\right)\cdot\mathbf{d},\label{Init_cond_general_matr}
\end{equation}
where
\begin{equation}
A=\left(-p_{S}G_{\Sigma}+p_{\Sigma}G_{S}\right)\Lambda,\qquad B=p_{S}G_{\Sigma}+p_{\Sigma}G_{S}.
\end{equation}


\section{Analysis of the skyrmion's equation of motion}

\label{Sec_Analysis}

The equation of motion for the ferrimagnetic skyrmion (\ref{Newton_eq_damped})
has a similarity with the equation of motion of a charge $q$ in the
magnetic field $\mathbf{B}$, i.e., $\left(\mathbf{G}_{S}-\mathbf{G}_{\Sigma}\right)\times\mathbf{V}\Leftrightarrow q\mathbf{V}\times\mathbf{B}$.
However, the time-dependent term $M\dot{\mathbf{U}}$ in Eq. (\ref{K_def})
makes the behavior of the skyrmion different in the presence of a
time-dependent force and/or spin current. In this section some solutions
of Eq. (\ref{Newton_eq_damped}) will be shown.

One can expect that the mass of the ferrimagnetic skyrmion goes to
zero in the ferromagnetic limit $c\rightarrow0$, because the mass
of the ferromagnetic skyrmion is zero. However, $M$ in Eq. (\ref{M_damped})
does not depend on spins and is a constant. What is the resolution
of this puzzle?

First, for the realistic model with disorder, the analytical approach
used above can be only valid if there are a sufficient number of RE
spins in the region of the TM skyrmion, $\pi\left(\lambda/a\right)^{2}c\gg1$.
For $\lambda_{\mathrm{eff}}/a=18$ for our single skyrmion in CoGd,
this translates into $c\gg10^{-3}$. If this condition is not satisfied,
there should be a large data scatter. The numerical calculations reported
below were performed down to $c=10^{-2}$, and the data scatter remained
moderate with no significant deviations from the analytical predictions
detected.

The idealized no-disorder model in which all RE spins have the length
$c\Sigma$ is self-consistent for any $c$, and it directly corresponds
to our analytical approach. For this model, in the limit $c\rightarrow0$,
the ferrimagnetic skyrmion does not show any inertia, despite $M$
being nonzero. That is, as the RE sublattice gradually disappears,
its action on the remaining TM sublattice disappears, too, than leads
to the conventional Thiele equation for the TM spins, as it should
be. Indeed, the interaction force $\mathbf{F}_{S\Sigma}$ given by
Eq. (\ref{FSSigma_res}) vanishes if one of the two spin densities
vanish, and in this case the two Thiele equations decouple.

As an example, one can create an initial displacement $\mathbf{d}$,
say, by a short pulse of a spin current, and then the skyrmion will
move ballistically in the absence of forces and spin currents, making
circles. What happens with this motion if $c\rightarrow0$? Note that
there are only $S-\Sigma$ cross-terms in the matrix in Eq. (\ref{Init_cond_general_matr}).
When the RE sublattice disappears, one must look exclusively at the
TM sublattice, setting $p_{S}=1$ and $p_{\Sigma}=0$. But in all
terms $p_{S}$ multiplies by $G_{\Sigma}$ that goes to zero, so the
whole expression for $\mathbf{V}$(with $\mathbf{U}=0$) vanishes
whatever is the displacement $\mathbf{d}$. As the initial condition
for Eq. (\ref{Newton_eq_damped}) with $\mathbf{K}=0$ is $\mathbf{V}(0)=0$,
the solution is $\mathbf{V}=0$ at all times, that is, there is no
ballistic motion.

More general, Eq. (\ref{Newton_eq_damped}) can be rewritten in terms
of the complex variable $\xi\equiv V_{x}+iV_{y}$ as
\begin{equation}
M\dot{\xi}-iM\tilde{\Omega}\xi=K_{x}+iK_{y},
\end{equation}
where
\begin{equation}
\tilde{\Omega}=\frac{G_{S}-G_{\Sigma}+i\left(\Gamma_{S}+\Gamma_{\Sigma}\right)}{M}=\Omega+\frac{i\Lambda\left(G_{S}+G_{\Sigma}\right)}{M}
\end{equation}
is the damped SRC frequency. For $\mathbf{K}=\mathbf{const}$ the
solution of this equation reads
\begin{equation}
\xi(t)=\xi(0)e^{i\tilde{\Omega}t}+\left(1-e^{i\tilde{\Omega}t}\right)\frac{iK_{x}-K_{y}}{M\tilde{\Omega}},\label{xi(t)_result_general}
\end{equation}
where $\xi(0)$ is the initial value. In the presence of damping the
exponentials vanish at large times, which leads to the asymptotic
solution $\xi(\infty)=\left(iK_{x}-K_{y}\right)/\left(M\tilde{\Omega}\right)$.
If the damping is small, the skyrmion is asymptotically moving in
the direction perpendicular to the vector $\mathbf{K}$, similarly
to the electric charge in the perpendicular electric and magnetic
fields. In the absence of damping for $\mathbf{K}=\mathbf{0}$, the
skyrmion is making a circle with the cyclotron radius
\begin{equation}
R_{\mathrm{cyc}}=\frac{\left|\xi\right|}{\Omega}=d\frac{p_{\Sigma}G_{S}+p_{S}G_{\Sigma}}{\left|G_{S}-G_{\Sigma}\right|}=d\frac{p_{\Sigma}S+p_{S}c\Sigma}{\left|S-c\Sigma\right|},\label{R_cyc}
\end{equation}
where Eq. (\ref{Init_cond_general_matr}) was used. As the TM and
RE skyrmions have different velocities, $R_{\mathrm{cyc}}$ depends
on the definition of the ferrimagnetic-skyrmion center via the weight
factors $p_{S}$ and $p_{\Sigma}$. However, the difference between
different definitions remains as small as $d$ even near the AMC point
where $R_{\mathrm{cyc}}$ diverges. It becomes clear if one rewrites
$R_{\mathrm{cyc}}$ using $p_{S}+p_{\Sigma}=1$ as
\begin{equation}
R_{\mathrm{cyc}}=d\frac{c\Sigma+p_{\Sigma}\left(S-c\Sigma\right)}{\left|S-c\Sigma\right|}.
\end{equation}
Here, the divergent part of $R_{\mathrm{cyc}}$ does not depend on
the definition of the skyrmion's center. In the limit $c\rightarrow0$,
one should focus on the TM skyrmion and set $p_{S}=1$ and $p_{\Sigma}=0$.
In this case, $R_{\mathrm{cyc}}\propto c$ and goes to zero for $c\rightarrow0$.
This is an expected result as there is no inertia effects and hence
no cyclotron resonance for the ferromagnetic skyrmion.

Another example is the dynamics due to the time-dependent spin current
applied in the absence of forces. In this case, one has $\mathbf{U}=p_{S}\mathbf{v}_{S}+p_{\Sigma}\mathbf{v}_{\Sigma}$,
and one can rewrite Eq. (\ref{Newton_eq_damped}) as
\begin{eqnarray}
M\left(\dot{\mathbf{V}}-p_{S}\dot{\mathbf{v}}_{S}-p_{\Sigma}\dot{\mathbf{v}}_{\Sigma}\right)\nonumber \\
-\mathbf{G}_{S}\times\left(\mathbf{V}-\mathbf{v}_{S}\right)+\mathbf{G}_{\Sigma}\times\left(\mathbf{V}-\mathbf{v}_{\Sigma}\right)\nonumber \\
+\Gamma_{S}\left(\mathbf{V}-\mathbf{v}_{S}\right)+\Gamma_{\Sigma}\left(\mathbf{V}-\mathbf{v}_{\Sigma}\right) & = & \mathbf{0}.
\end{eqnarray}
If the RE sublattice disappears, $\mathbf{G}_{\Sigma}\rightarrow0$
and $\Gamma_{\Sigma}\rightarrow0$, then one sets $p_{S}=1$ and $p_{\Sigma}=0$,
and the equation becomes
\begin{equation}
M\left(\dot{\mathbf{V}}-\dot{\mathbf{v}}_{S}\right)-\mathbf{G}_{S}\times\left(\mathbf{V}-\mathbf{v}_{S}\right)+\Gamma_{S}\left(\mathbf{V}-\mathbf{v}_{S}\right)=\mathbf{0}.
\end{equation}
The obvious solution is $\mathbf{V}(t)=\mathbf{v}_{S}(t)$, same as
the solution for the ferromagnetic skyrmion, describing transporting
the TM skyrmion by the spin current. In this case there is no inertia
as well. The bottom line is that in our analytical model the mass
of the ferrimagnetic skyrmion does not go to zero but becomes hidden
in the ferromagnetic limit $c\rightarrow0$.

In the general case, the linear response to the oscillating spin current
$\mathbf{v}_{S}(t)=v_{S,0}\mathbf{e}_{x}\sin\left(\omega t\right)$
and $\mathbf{v}_{\Sigma}(t)=0$, calculated using Eq. (\ref{Newton_eq_damped})
has the form
\begin{equation}
V_{x}+iV_{y}=A\sin\left(\omega t\right)+B\cos\left(\omega t\right),
\end{equation}
where
\begin{equation}
A=v_{S,0}\left[1-\frac{\left(\tilde{\Gamma}-i\Omega\right)\left(\tilde{\Gamma}_{\Sigma}+i\Omega_{\Sigma}\right)}{\omega^{2}-\Omega^{2}-2i\Omega\tilde{\Gamma}+\tilde{\Gamma}^{2}}\right]
\end{equation}
and
\begin{equation}
B=v_{S,0}\frac{\omega\left(\tilde{\Gamma}_{\Sigma}+i\Omega_{\Sigma}\right)}{\omega^{2}-\Omega^{2}-2i\Omega\tilde{\Gamma}+\tilde{\Gamma}^{2}}.
\end{equation}
Here $\Omega_{\Sigma}\equiv G_{\Sigma}/M$ and
\begin{equation}
\tilde{\Gamma}\equiv\left(\Gamma_{S}+\Gamma_{\Sigma}\right)/M,\qquad\tilde{\Gamma}_{\Sigma}\equiv\Gamma_{\Sigma}/M
\end{equation}
are relaxation frequencies. In the limit $c\rightarrow0$, one has
$\Omega_{\Sigma}\rightarrow0$ and $\tilde{\Gamma}_{\Sigma}\rightarrow0$,
so that the terms describing the cyclotron motion vanish and the dynamics
becomes trivial: $V_{x}(t)=v_{S,0}\sin\left(\omega t\right)=v_{S}(t)$
and $V_{y}(t)=0$. If the RE subsystem is present, the solution above
describes the cyclotron resonance at the frequency $\omega\cong\varOmega$,
where $\Omega$ is given by Eq. (\ref{Omega_cyclotron}). In this
example, the terms with the time derivatives in $\mathbf{K}$ given
by Eq. (\ref{K_def}) are crucial.

\section{Numerical methods}

\label{Sec_Numerical-methods}

The numerical procedures employed here are (i) energy minimization
at $T=0$ and (ii) solving the undamped Landau-Lifshitz equation (\ref{LL})
for our many-spin model. The energy minimization \citep{garchupro2013}
combines aligning the spin $\mathbf{s}_{i}$ with its effective field
$\mathbf{H}_{\mathrm{eff},i}=-\partial\mathcal{H}/\partial\mathbf{S}_{i}$
with the probability $\eta$ and flipping the spin around the effective
field, $\mathbf{S}_{i}\Rightarrow2\left(\mathbf{S}_{i}\cdot\mathbf{H}_{\mathrm{eff},i}\right)\mathbf{H}_{\mathrm{eff},i}/H_{\mathrm{eff},i}^{2}-\mathbf{S}_{i}$
with the probability $1-\eta$ (the so-called overrelaxation). The
algorithm uses vectorized updates of columns of spins in checkered
sublattices that allows parallelization of the computation. For the
different-site interactions, such as exchange and DMI, overrelaxation
is a kind of conservative pseudodynamics, which allows for better
exploration of the phase space of the system. Choosing a small value
of $\eta$ that has a meaning of the damping in this procedure (the
main choice being $\eta=0.03$) allows achieving a much faster and
deeper energy minimization then just the field alignment, $\eta=1$.
For the model with single-site interactions, such as the uniaxial
anisotropy, overrelaxation leads to the energy decrease (see Eq. (8)
of Ref. \citep{gar2025jpcm}), so that one can use $\eta=0$ (which
turns to be much more efficient than $\eta=1$). Still we use $\eta=0.03$
in all cases for the uniformity of the approach. This method of energy
minimization is much faster then solving the damped LL equation implemented
in micromagnetic packages.

\begin{figure}
\begin{centering}
\includegraphics[width=8cm]{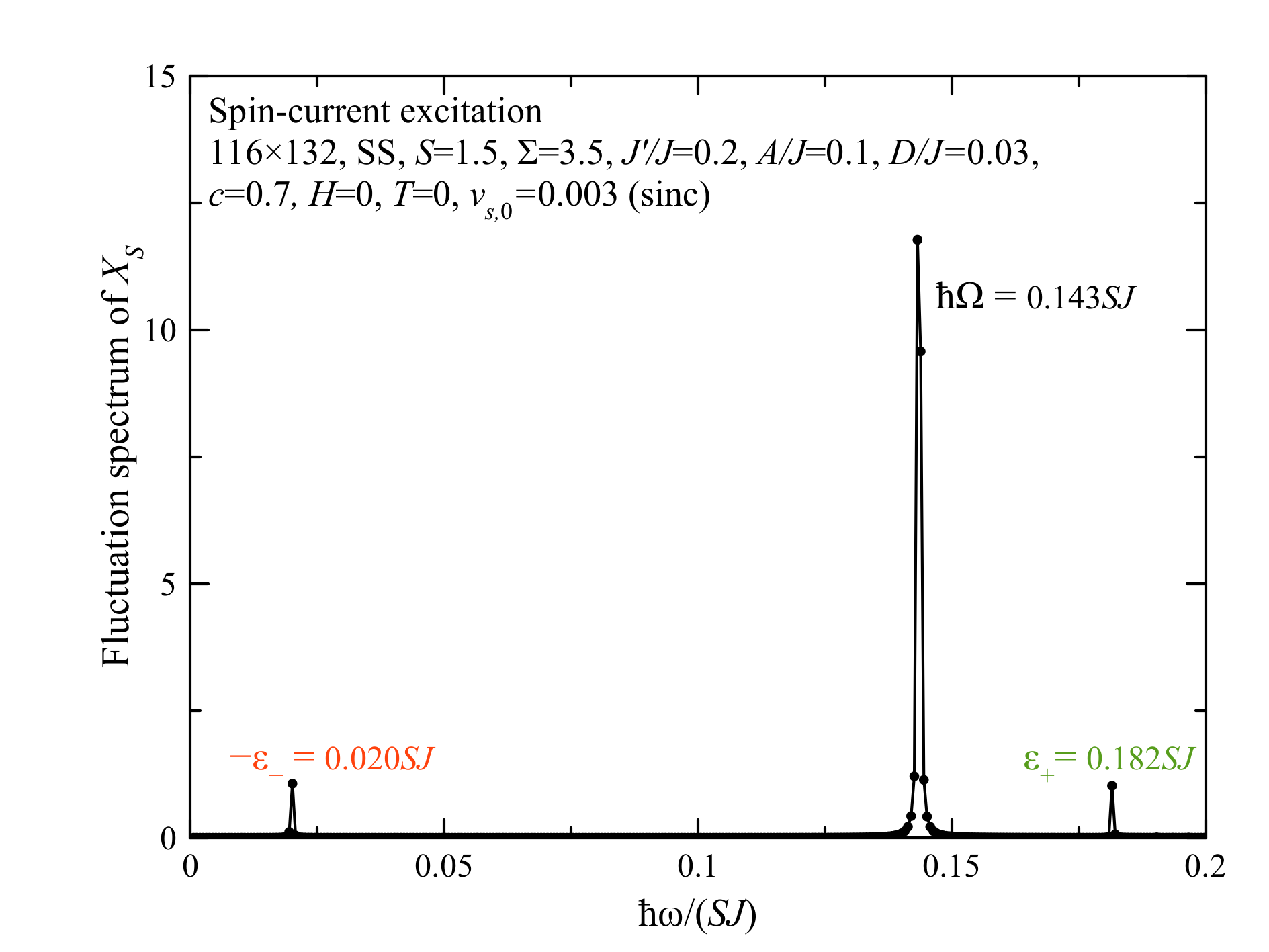}
\par\end{centering}
\caption{Fluctuation spectrum of the TM skyrmion position, $X_{S}$, in the
$116\times132$ ferrimagnetic system with the RE concentration $c=0.7$
excited by a sinc spin current. The biggest peak corresponds to the
skyrmion cyclotron resonance, while two small peaks correspond to
the ferrimagnetic $k=0$ spin-wave modes $\varepsilon_{\pm}$ shown
in Fig. \ref{Fig_eps_pm_vs_c_analytical}. }\label{Fig_FSX}
\end{figure}
\begin{figure}
\begin{centering}
\includegraphics[width=8cm]{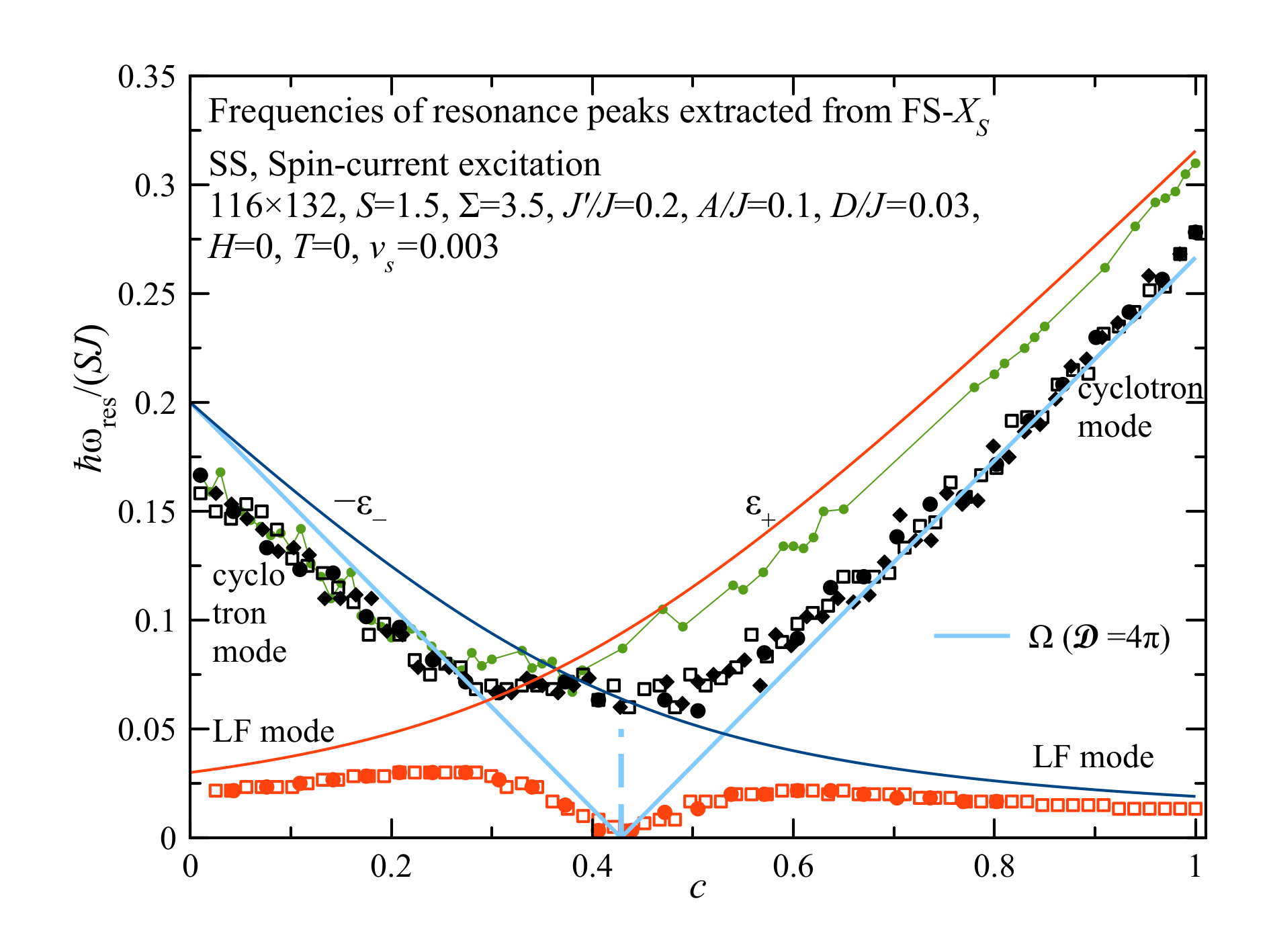}
\par\end{centering}
\begin{centering}
\includegraphics[width=8cm]{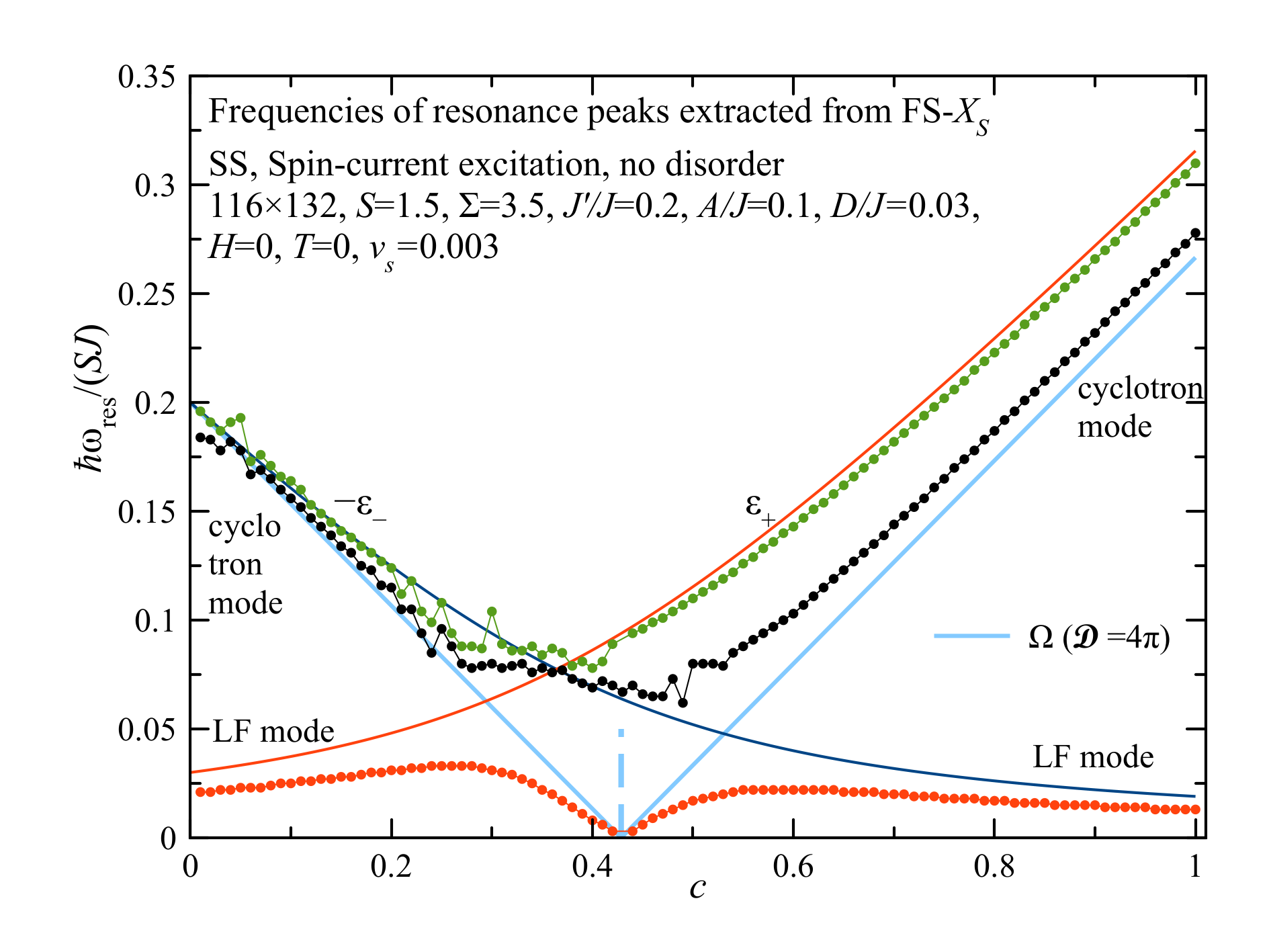}
\par\end{centering}
\caption{The RE-concentration dependence of the frequencies of the excitation
modes in the model of CoGd ferrimagnet. The system containing a single
skyrmion in a $116\times132$ system was excited by the sinc spin
current, and the fluctuation spectrum of the TM skyrmion position,
$X_{S}$, was observed. The theoretical result for the SCR frequency
(\ref{Omega_cyclotron-BP}) is shown by the light-blue line. Uniform
ferrimagnetic modes $\varepsilon_{\pm}$ are also shown. Top: the
realistic model with disorder; Bottom: the idealized no-disorder model.
}\label{Fig_omega_res}
\end{figure}
\begin{figure}
\begin{centering}
\includegraphics[width=8cm]{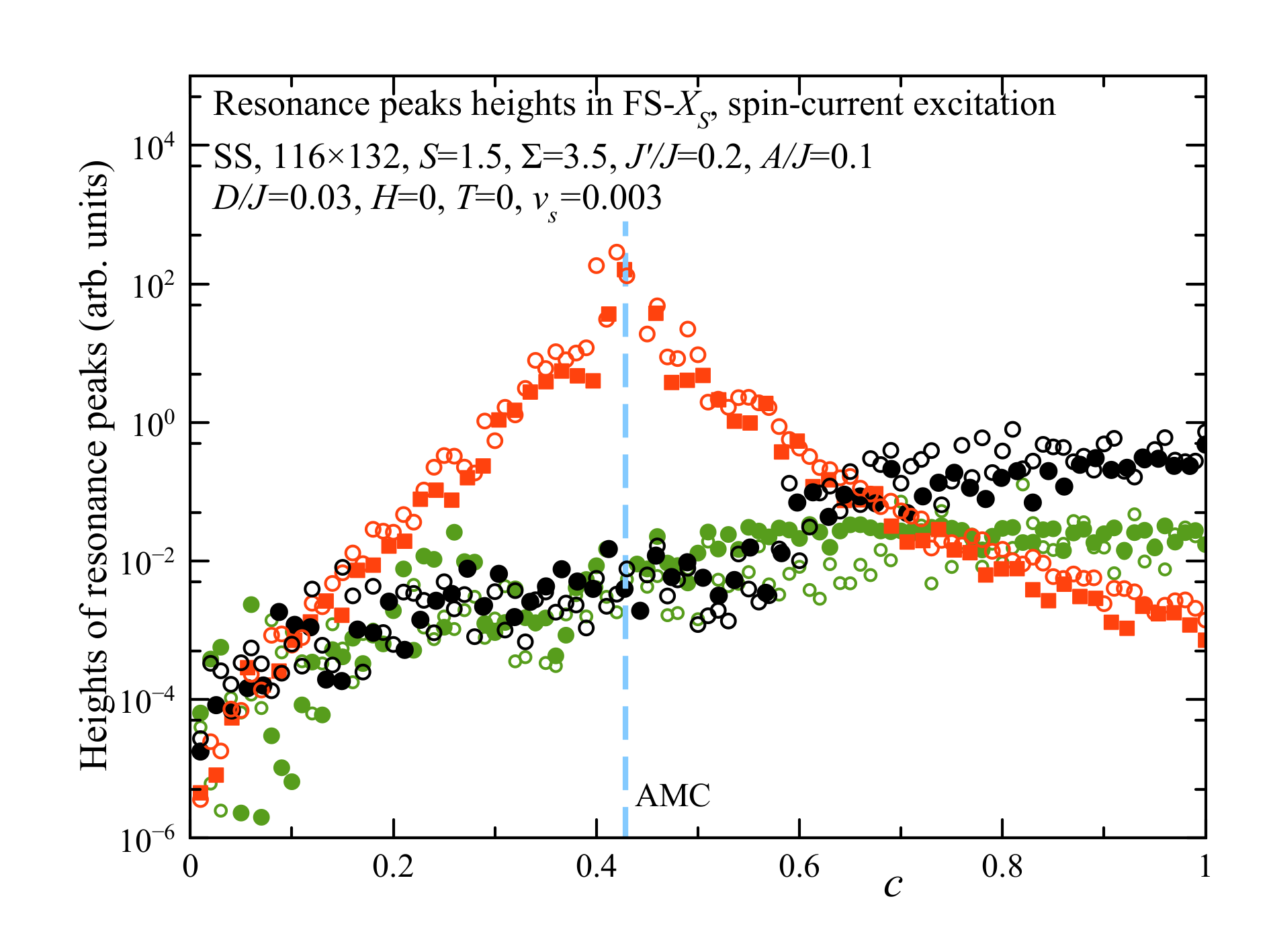}
\par\end{centering}
\caption{Heights of resonance peaks in the FS in Fig. \ref{Fig_omega_res}.
}\label{Fig_Peak_heights}
\end{figure}
\begin{figure}
\begin{centering}
\includegraphics[width=8cm]{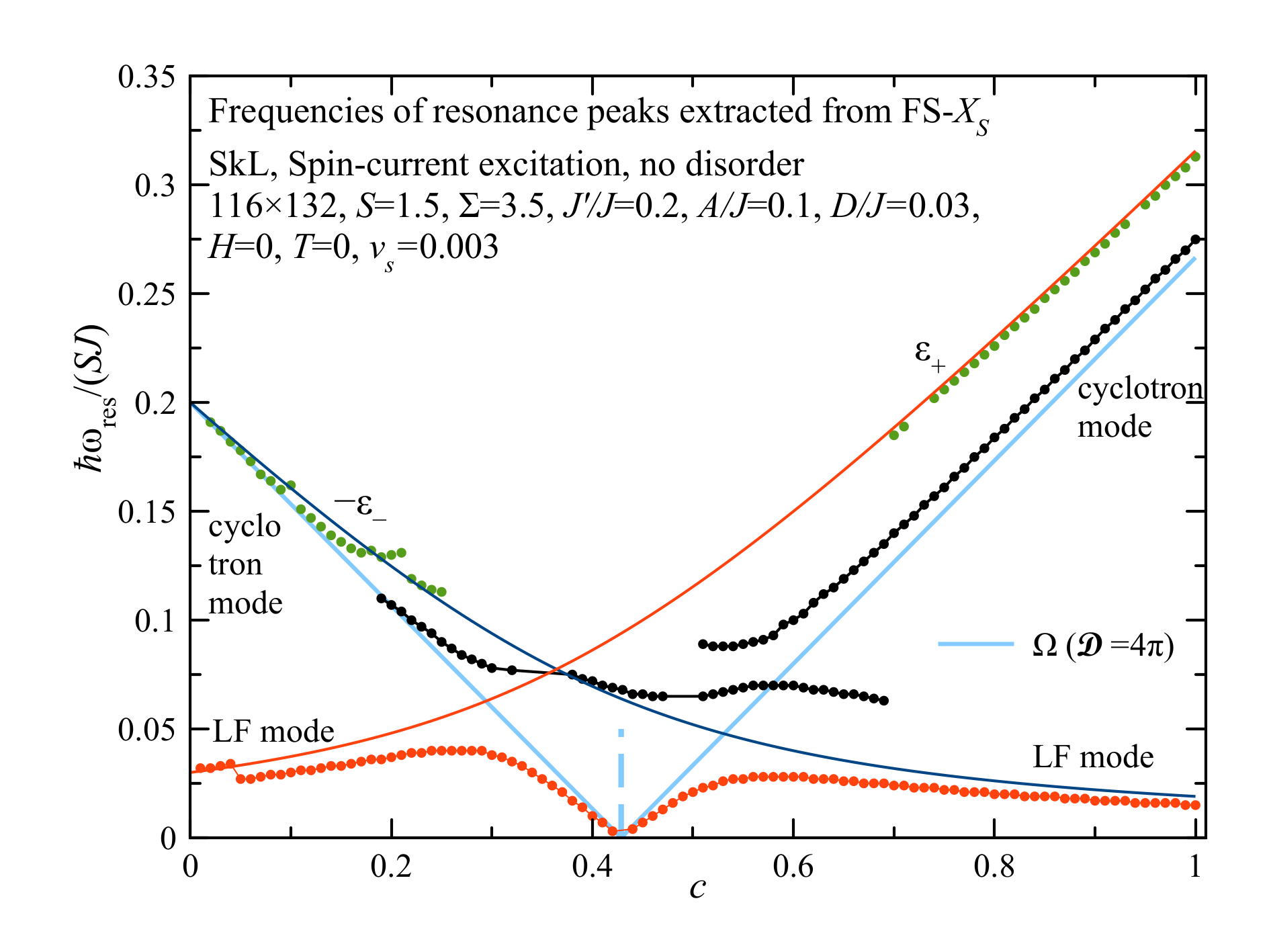}
\par\end{centering}
\begin{centering}
\includegraphics[width=8cm]{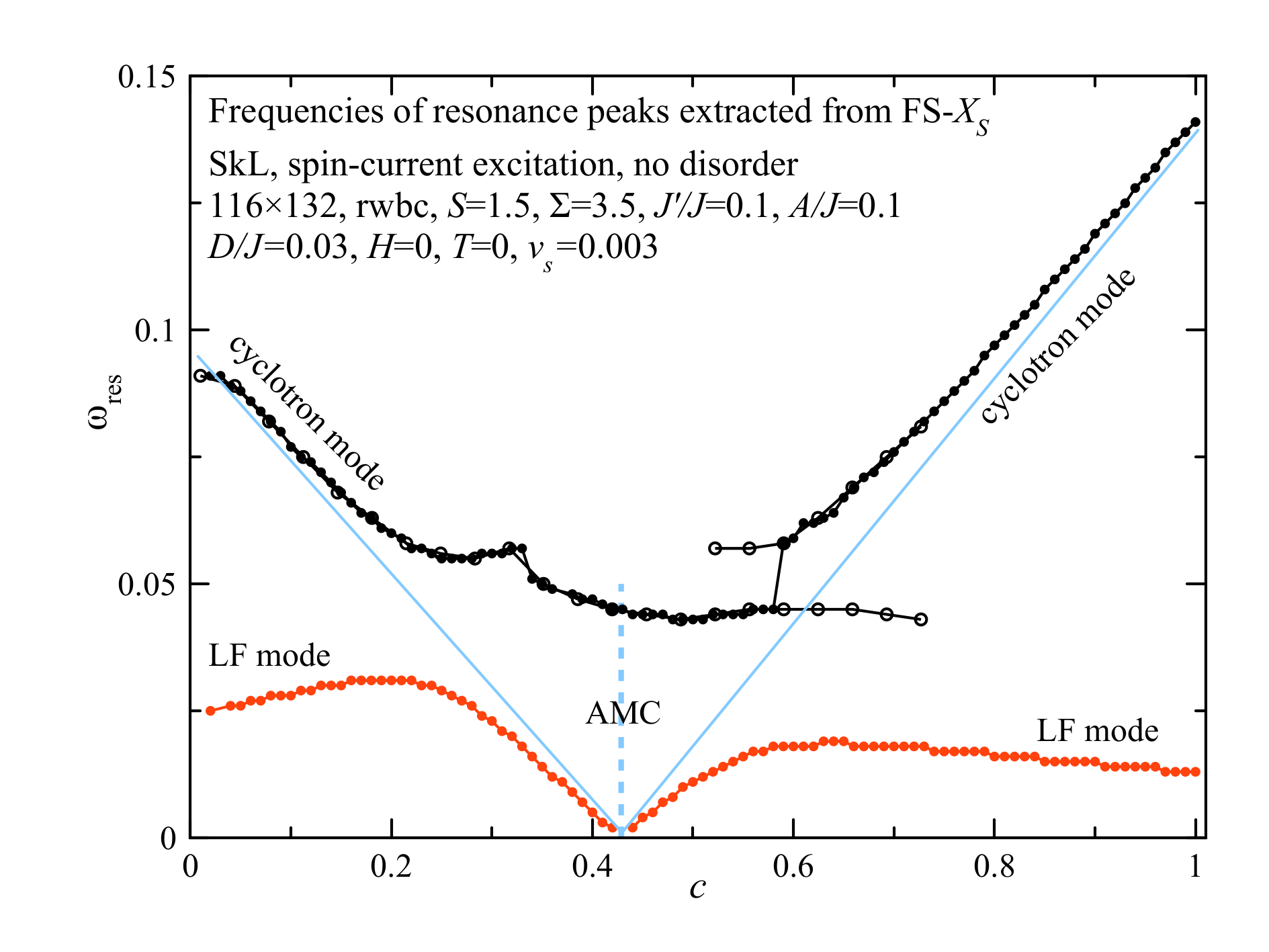}
\par\end{centering}
\caption{The RE-concentration dependence of the frequencies of the excitation
modes in the model of CoGd ferrimagnet. The $116\times132$ no-disorder
system with SkL was excited by the sinc spin current, and the fluctuation
spectrum of the TM skyrmion position, $X_{S}$, was observed. Top:
$J'/J=0.2$ (the standard choice). Bottom: $J'/J=0.1$.}\label{Fig_omega_res_vs_c_SkL}
\end{figure}

To compute the dynamical evolution of the system, we employed the
fifth-order Butcher's Runge-Kutta ordinary differential equation (ODE)
solver, RK5, which makes six function evaluations per integration
step (see, e.g., the Appendix in Ref. \citep{gar17prb}). It is much
more precise than the classical RK4 solver. Precision in dynamical
computation is important, since numerical errors accumulate over a
large evolution period, leading to the energy drift in conservative
systems under consideration. If the computation is long, as the computation
of the spectrum of the absorbed power using the fluctuation-dissipation
theorem (FDT) \citep{garchu2022prb,DG-EC-PRB2025,DG-EC-PRB2026},
one needs to perform the energy correction \citep{gar21pre} from
time to time. 

In this paper, only the excitation spectrum of the system at $T=0$
is investigated, and this can be done much faster if one computes
the fluctuation spectrum (FS) of the dynamical quantity of interest
$F(t)$. The FS is defined by
\begin{equation}
\mathrm{FS}(\omega)=\frac{1}{t_{\max}}\left|\intop_{0}^{^{t_{\max}}}F(t)e^{i\omega t}dt\right|^{2}.\label{FS}
\end{equation}
 In this case, $t_{\max}$ does not need to be very long, as in the
FDT computations, thus one does not need to perform the energy correction.
To initiate the dynamics, one could create a deviation of $F$ from
its equilibrium value with Monte Carlo at a low temperature $T$.
However, there is a much more elegant and efficient method using the
sinc excitation of the system. In the case of the excitation by the
spin current it is $v_{S}(t)=v_{S,0}\sin\left(\omega_{\max}t\right)/\left(\omega_{\max}t\right)$.
The Fourier spectrum of this function is a constant up to the cutoff
frequency $\omega_{\max}$ and zero above it, so that it excites all
modes in the interval $0<\omega<\omega_{\max}$. As the quantity $F$
we used either the $X$ displacement of the skyrmion (or the average
displacement of skyrmions in the SkL) in the TM subsystem, $X_{S}$,
from its equilibrium position or the $x$ component of the normalized
magnetic moment per site
\begin{equation}
\boldsymbol{\mu}=\frac{1}{\mathcal{N}}\sum_{i}\left(\mathbf{S}_{i}+\frac{g'}{g}p_{i}\mathbf{\Sigma}_{i}\right).\label{mu_vec_def}
\end{equation}
The modes' frequencies were searched for as the resonances in $\mathrm{FS}(\omega)$.
One dynamical computation provided the results for the excitations
at all frequencies within the interval of interest. We used the system
size $116\times132$ in lattice units (with periodic boundary conditions)
that neatly comprises a skyrmion lattice (SkL) of 12 skyrmions, as
shown in Fig. \ref{Fig_SkL}. This system size is sufficient to find
the excitation spectrum of the system in the cases of a single skyrmion
(SS) and the SkL. We generated a random distribution of RE atoms with
the concentration $c$. 

Also, we numerically studied the \textit{no-disorder model} in which
all RE spins were assigned the length $c\Sigma$. This model uses
the same simplification as our analytical approach, in which disorder
effects are ignored and the RE concentration $c$ enters only in the
combination $c\Sigma$. This model does not show the scatter due to
the disorder in the realistic model. As a justification of the no-disorder
model, one can mention that in the presence of disorder there are
localized excitation modes \citep{garchu2023prb}. For uniform ferrimagnets,
this leads to the splitting of the peaks in the FS into many subpeaks,
as shown in Fig. 9 of Ref. \citep{DG-EC-PRB2026}. This is the source
of the data scatter for our $116\times132$ model if the original
model with disorder is used. For large systems, there are many subpeaks
that merge and form a broadened line, see Fig. 10 of Ref. \citep{DG-EC-PRB2026}.
In this case, the data scatter disappears. However, performing computations
for larger systems for many values of $c$ requires too much computing
time. The no-disorder model is an approximation that neglects the
effect of localized modes and line broadening, allowing to obtain
smooth results for moderate-size systems.

The position of the TM skyrmion's center $\mathbf{R}$ was defined
with the help of the skyrmion-locator formula \citep{garchu24jmmm}
\begin{equation}
\mathbf{R}=\left.\sum_{S_{j,z}>0}\mathbf{r}_{j}S_{j,z}^{2}\right/\sum_{S_{j,z}>0}S_{j,z}^{2},\label{Skyrmion_locator}
\end{equation}
where $j\in\mathrm{Skyrmion}$ are all lattice sites that belong to
the skyrmion. Here the weight factor $S_{z,j}^{2}$ favors the sites
closer to the skyrmion's top. The same formula was used for the SkL
to describe the motion of all skyrmions.

The skyrmion size $\lambda_{\mathrm{eff}}$ was computed as \citep{caichugar2012prb}
\begin{equation}
\lambda_{\mathrm{eff}}^{2}=\frac{n-1}{2^{n}\pi N_{S}}\sum_{i}\left(1+\frac{S_{i,z}}{S}\right)^{n}\label{lam_eff_def}
\end{equation}
for the TM skyrmion and a similar formula for the RE skyrmion. Here
$N_{S}$ is the number of skyrmions, in our case $N_{S}=\left|Q\right|$.
This formula yields the exact skyrmion size $\lambda$ for the Belavin-Polyakov
pure-exchange skyrmion for any $n$. We use $n=4$ to effectively
cut the skyrmion's tails and focus on the skyrmion core. For our CoGd
parameters set, the skyrmion size is $\lambda_{\mathrm{eff}}/a\approx18.4$
for SS and $\lambda_{\mathrm{eff}}/a\approx8.9$ for the SkL, see
Fig. \ref{Fig_SkL}.

Computations were performed using our own compiled, vectorized, and
parallelized codes written in Wolfram Mathematica. For the number
crunching we employed prosumer-grade computers available to our group.
In the computations, we used moderate integration times such as $t_{\max}\approx6000$
with $SJ\Rightarrow1$ and $\hbar\Rightarrow1$, as usual. As the
system size was not large and the computation times were not long,
the numerical part of the work was not tough, given that we adopted
the codes used in our preceding projects.

\section{Numerical results}

\label{Sec_Numerical-results}

\begin{figure}
\begin{centering}
\includegraphics[width=8cm]{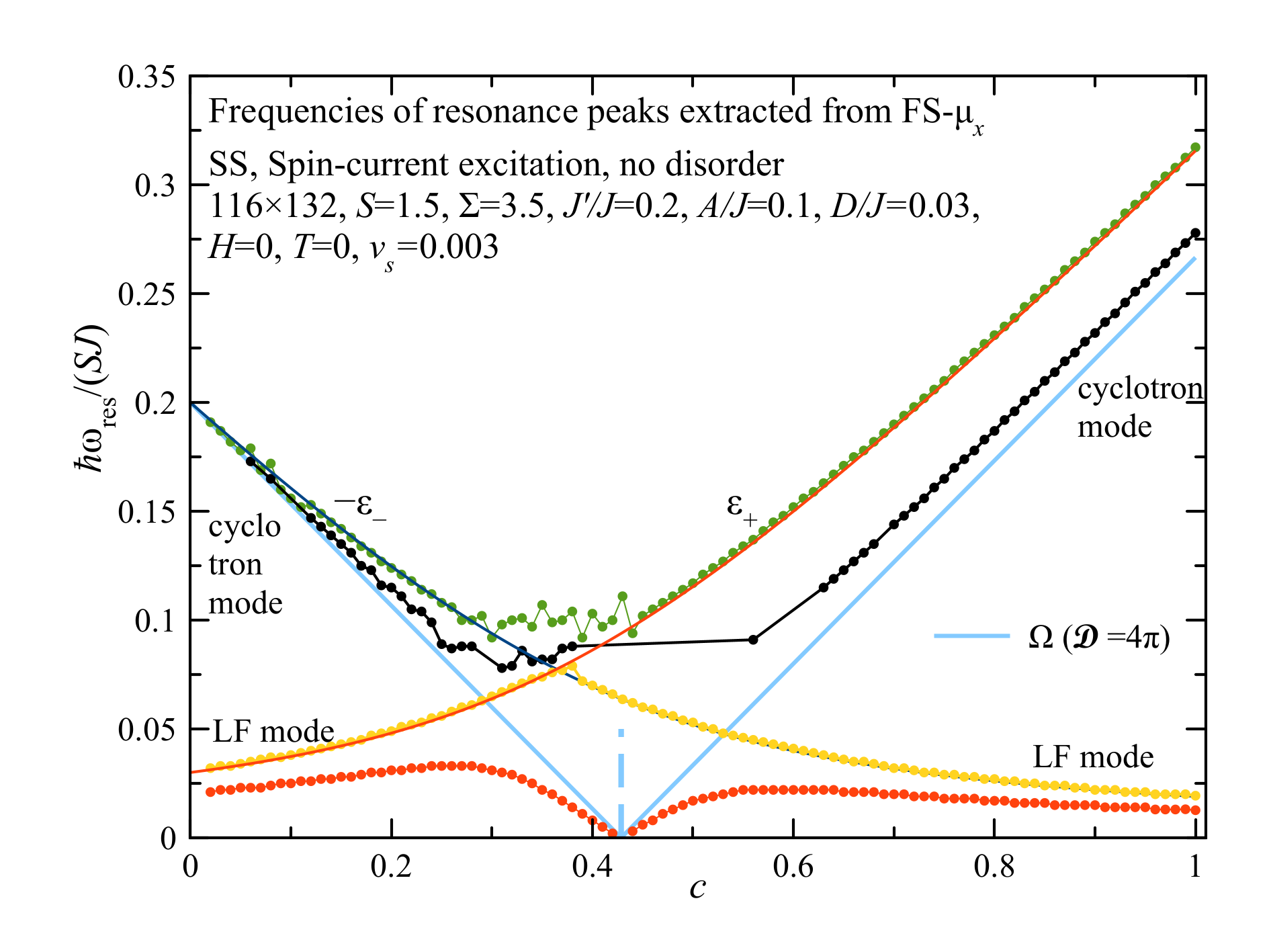}
\par\end{centering}
\begin{centering}
\includegraphics[width=8cm]{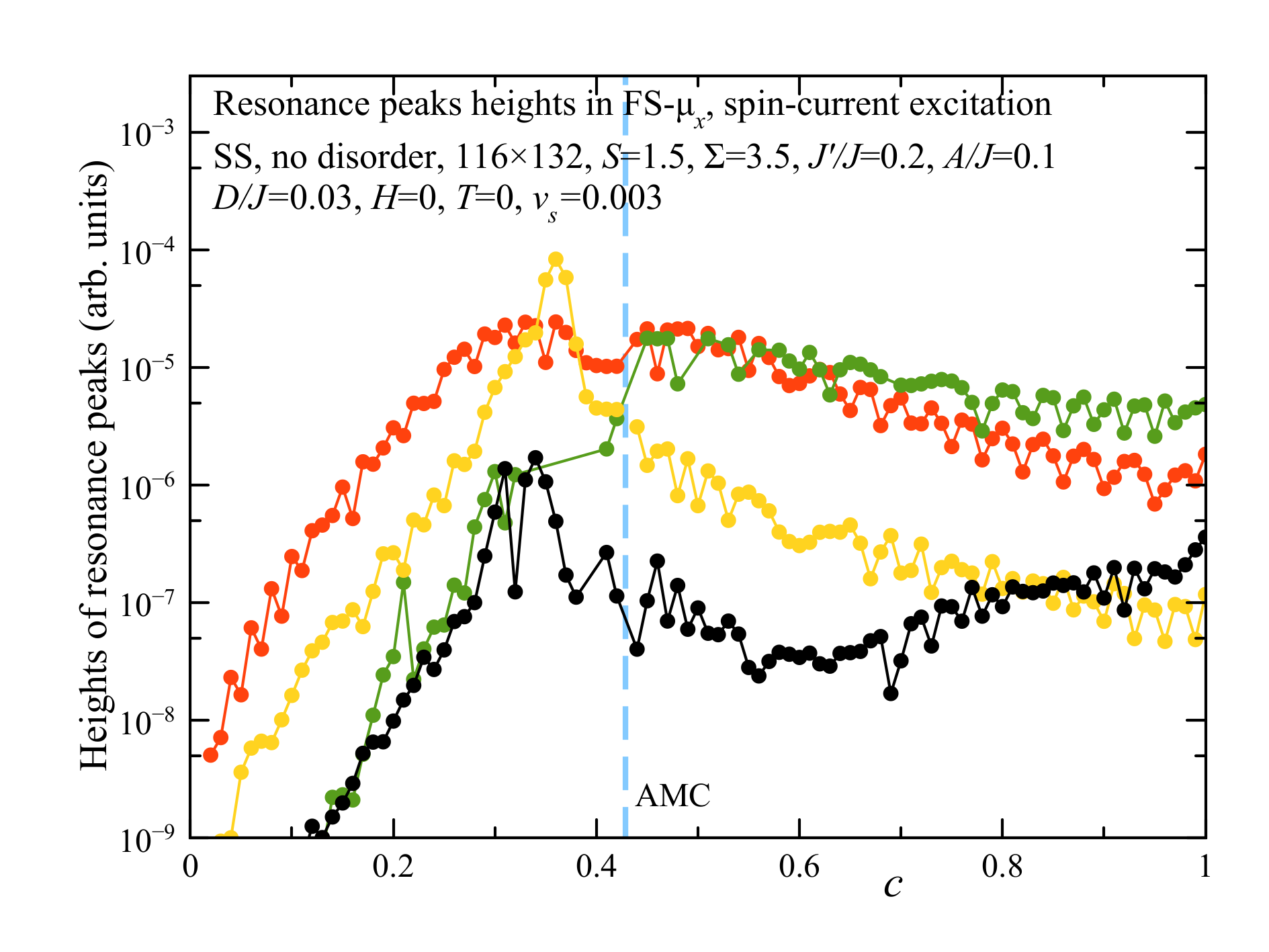}
\par\end{centering}
\caption{Top: The RE-concentration dependence of the frequencies of the excitation
modes in the model of CoGd ferrimagnet. The system containing a single
skyrmion in a $116\times132$ system was excited by the sinc spin
current, and the fluctuation spectrum of the transverse magnetization
component, $\mu_{x}$, was observed. The theoretical result for the
SCR frequency (\ref{Omega_cyclotron-BP}) is shown by the light-blue
line. Uniform ferrimagnetic modes $\varepsilon_{\pm}$ are also shown.
Bottom: the heights of the peaks in the top panel. }\label{Fig_omega_res_vs_c_116x132_spin_curr_mux_no_disorder}
\end{figure}
\begin{figure}
\centering{}\includegraphics[width=8cm]{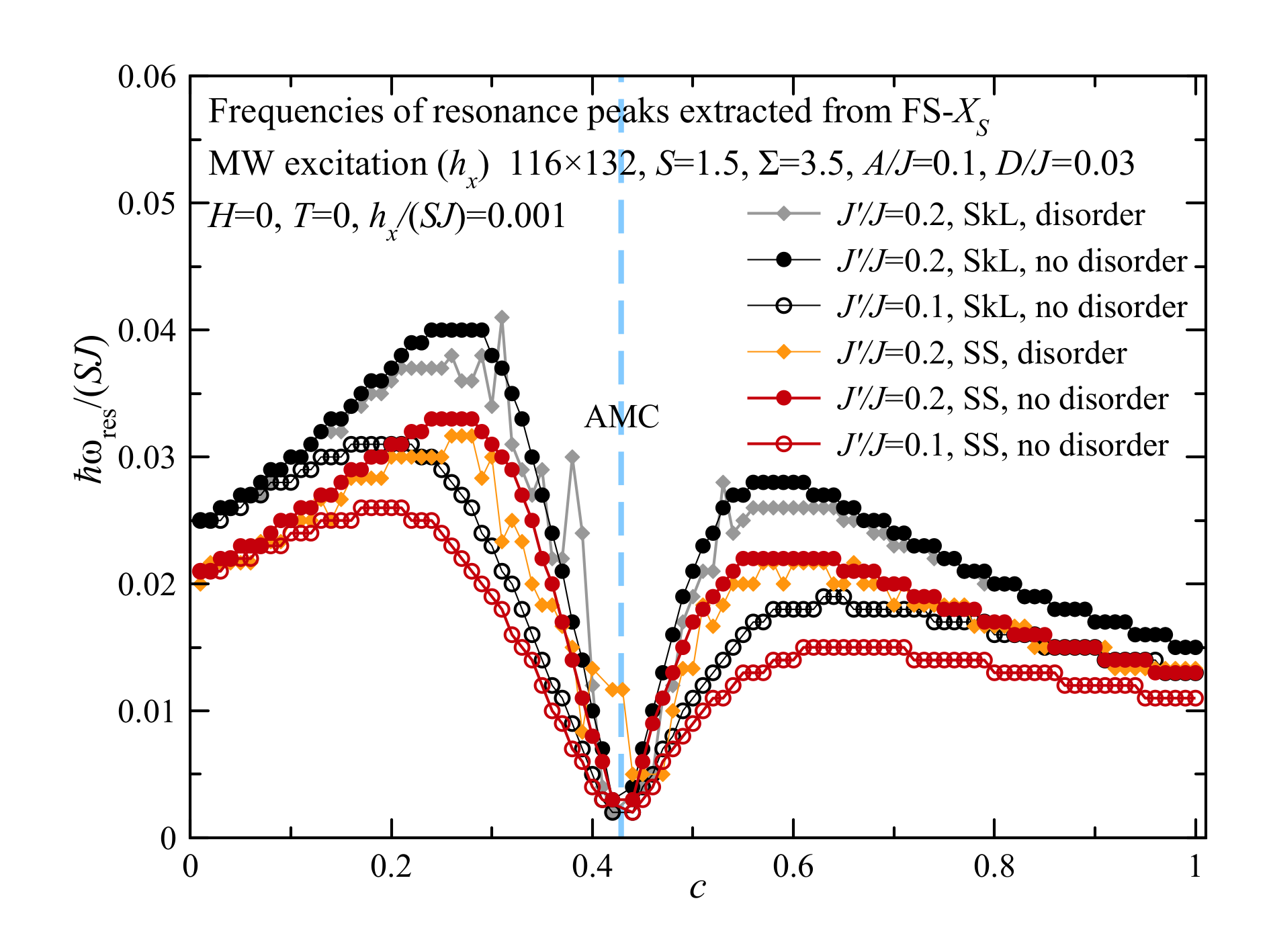}\caption{The RE-concentration dependence of the lower-mode frequenciy in the
model of CoGd ferrimagnet. The $116\times132$ system was excited
by the sinc microwaves, and the fluctuation spectrum of the TM skyrmion
displacement, $X_{S}$, was observed. The results are presented for
a single skyrmion (SS) and the SkL, with and without disorder, and
for the intersublattice coupling $J'/J=0.2$ and 0.1. }\label{Fig_omega_res_vs_c_MW_XS_lower_mode}
\end{figure}

\begin{figure}
\begin{centering}
\includegraphics[width=8cm]{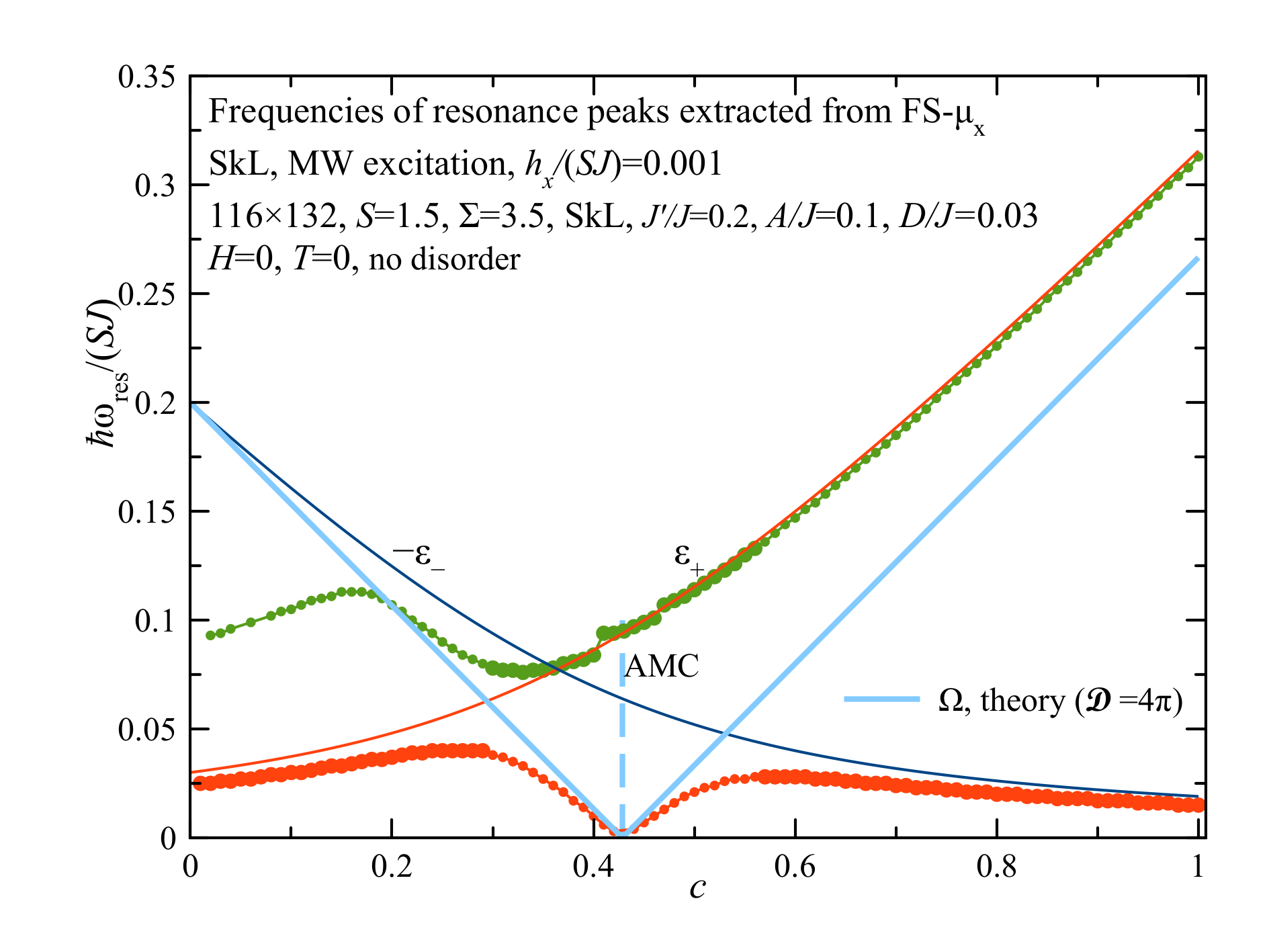}
\par\end{centering}
\begin{centering}
\includegraphics[width=8cm]{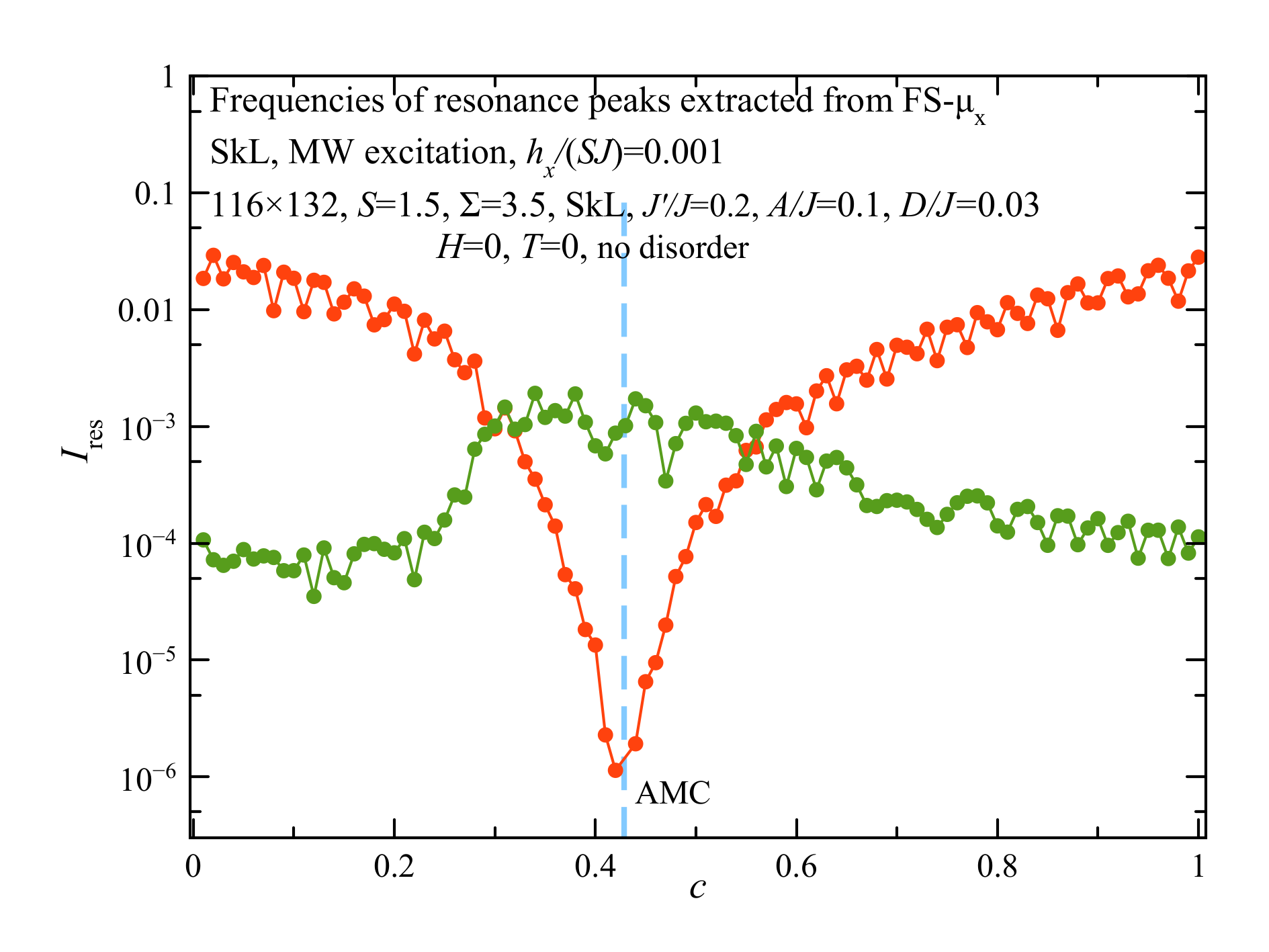}
\par\end{centering}
\caption{Top: The RE-concentration dependence of the frequencies of the excitation
modes in the model of CoGd ferrimagnet. The $116\times132$ system
with SkL was excited by the sinc microwaves, and the fluctuation spectrum
of the transverse magnetization component, $\mu_{x}$, was observed.
The regions, where the modes are strong, are shown by bigger markers.
Bottom: the heights of the peaks in the top panel.}\label{Fig_omega_res_vs_c_SkL_MW_mux}
\end{figure}

The main numerical experiment was exciting the system at $T=0$ with
the sinc spin current and extracting the frequencies of the excitation
modes from the fluctuation spectrum of the skyrmion's displacement,
defined by Eq. (\ref{Skyrmion_locator}). Computations were done on
the system containing a single skyrmion and the skyrmion lattice,
obtained by the energy minimization. It turned out that the cyclotron
motion of the skyrmion is coupled to the magnetization precession
in the system. Thus, one can also excite the system by microwaves
and extract the modes' frequencies from the FS of the in-plane components
of the magnetization, Eq. (\ref{mu_vec_def}). This gives four different
numerical experiments altogether. On top of it, one can experiment
with a single skyrmion and the skyrmion lattice.

Figure \ref{Fig_FSX} shows the fluctuation spectrum of the TM skyrmion
displacement $X_{S}$ of a single skyrmion excited by the sinc spin
current for the RE concentration $c=0.7$. There are three peaks,
which can be identified as the $k=0$ spin-wave modes $\varepsilon_{\pm}$,
Eq. (\ref{epsilon_pm-general}), and the skyrmion cyclotron mode with
the frequency $\Omega$, Eq. (\ref{Omega_cyclotron-BP}). 

Figure \ref{Fig_omega_res} shows the RE-concentration dependence
of the frequencies of the excitation modes for the realistic model
with disorder (top) and the idealized no-disorder model (bottom).
In this figure, one can see the two ferrimagnetic modes $\varepsilon_{\pm}$
and the SCR mode. However, the modes are hybridized, forming a gap.
There is the upper part of the cyclotron mode (black) and its lower
part (red) that is strongly hybridized with the LF ferromagnetic mode.
The frequency of this lower mode goes to zero at the AMC point, as
predicted by Eq. (\ref{Omega_cyclotron}). The results for the no-disorder
model are smoother, as expected. The $c$-dependence of the peak heights
in Fig. \ref{Fig_Peak_heights} shows that the lower mode becomes
strong in the vicinity of the AMC point. This is expected because
here the cyclotron radius of the skyrmion diverges. For $c\rightarrow0$
all FS peak heights become small. The results for the model with disorder
are more scattered, and at some values of $c$ the data for the modes'
frequencies are missing. Also disorder decreases the frequencies,
especially those of the HF and SCR modes at small $c$.

For the CoGd parameters, the numerically computed value of $\mathcal{D}$,
Eq. (\ref{Dcal_def}) is $\mathcal{D}\simeq20.2$ that substantially
exceeds the BP value $\mathcal{D}=4\pi\simeq12.6$. The reason for
a larger $\mathcal{D}$ is that in the presence of the anisotropy
the skyrmion shape becomes flat-topped with a wall separating the
interior and exterior of the magnetic bubble. Large gradients inside
the wall increase $\mathcal{D}$. The theoretical SCR line with the
numerically computed $\mathcal{D}$ in Fig. \ref{Fig_omega_res} substantially
deviates from the numerical results, so it is not shown. In the contrary,
the theoretical SCR line with $\mathcal{D}=4\pi$ is in a good agreement
with the numerical data and is shown by the light-blue line in Fig.
\ref{Fig_omega_res} and other figures. This suggests that the universal
Eq. (\ref{Omega_cyclotron-BP}) for the SCR frequency is for some
reason better than Eq. (\ref{Omega_cyclotron}) in which $\Omega$
depends of other model parameters via the non-universal quantity $\mathcal{D}$.

Similar results were obtained for the skyrmion lattice, see Fig. \ref{Fig_omega_res_vs_c_SkL}.
Even for the no-disorder model, the results are less regular than
for a single skyrmion, and for some $c$-values some FS peaks could
not be found. The top panel of this figure shows the results for our
standard choice $J'/J=0.2$, whereas the lower panel shows the results
for a smaller value of the intersublattice coupling $J'/J=0.1$. In
the latter case, the cyclotron frequency far from the AMC point is
exactly twice as lower, in accordance with Eq. (\ref{Omega_cyclotron}).

Observing the fluctuation spectrum of the skyrmion's displacement
might be difficult because in the linear regime the displacement is
very small. However, since the cyclotron motion is coupled to the
ferrimagnetic modes, one also can extract the modes' frequencies from
the FS of the transverse magnetization, as shown in Fig. \ref{Fig_omega_res_vs_c_116x132_spin_curr_mux_no_disorder}.
Since the process is indirect, FS peaks are less well defined and
it is more difficult to extract the excitation frequencies. For this
reason, we show only the results for the no-disorder model. Here one
can see the same modes as in Fig. \ref{Fig_omega_res}, plus the unperturbed
LF ferrimagnetic branch (yellow). Again, the lower part of the spectrum
(red) is overall the strongest, although at $c\approx1$ the HF ferrimagnetic
mode (green) is the strongest.

Summarizing the results presented above, the most interesting of the
excitation modes, the lower mode, is robust, especially around the
angular-momentum compensation point where its frequency goes to zero.

Because of the coupling of the SCR mode to the ferrimagnetic resonance,
the former can also be excited by the microwaves. Figure \ref{Fig_omega_res_vs_c_MW_XS_lower_mode}
shows the RE concentration dependence of the frequency of the lower
mode in the model of CoGd ferrimagnet excited by the sinc MW. Here
the FS of the TM skyrmion displacement $X_{S}$ was used to obtain
the results for a single skyrmion (SS) and the SkL, with and without
disorder, and for the intersublattice coupling $J'/J=0.2$ and 0.1.
In all cases, the frequency goes to zero at the AMC point. For the
SkL the frequency is somewhat higher than for a SS. For the lower
$J$', the depression around the AMC point is broader, as expected.
Finally, the realistic system with disorder shows the scatter that
is absent in the idealized no-disorder model.

Figure \ref{Fig_omega_res_vs_c_SkL_MW_mux} shows the excitation modes'
frequencies and the FS peak heights for a $116\times132$ system with
SkL no disorder in the case when the system is excited by a sinc MW
and the FS of the transverse magnetization component $\mu_{x}$ is
used to extract the frequencies. In this case, the SCR mode was detected
only in the region around the AMC point, whereas the ferrimagnetic
modes $\varepsilon_{\pm}$ are seen at all $c$. Near the AMC point,
the LF mode hybridizes with the SCR and its frequency goes to zero.
However, the height of the FS peak corresponding to this mode becomes
small near the AMC point, because here the SCR is dominating in the
mixture of the modes, and the contribution of the magnetization precession
becomes small. The regions, where the modes are strong, are shown
by bigger markers in the top panel, while the concentration dependence
of the peaks' heights is shown in the bottom panel.

\begin{figure}
\begin{centering}
\includegraphics[width=8cm]{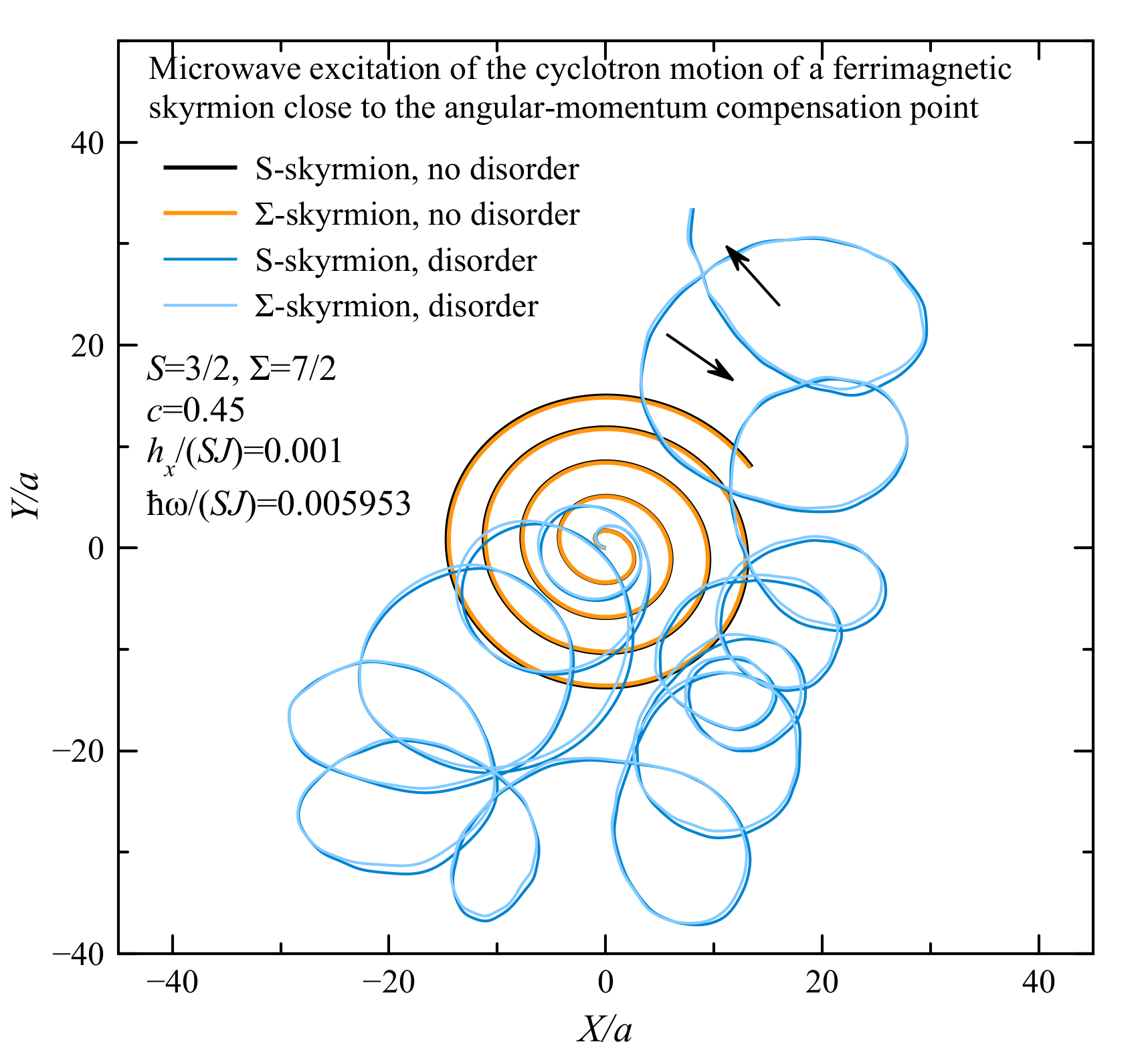}
\par\end{centering}
\caption{Skyrmion trajectory under MW excitation close to the AMC point.}\label{Fig_Skyrmion_Traj_MWx-excitation}

\end{figure}
Figure \ref{Fig_Skyrmion_Traj_MWx-excitation} illustrates the dynamics
of the excitation of the cyclotron mode close to the angular-momentum
compensation point, $c=0.45$ by the microwave field with the amplitude
$h_{x}/\left(SJ\right)=0.001$ at the frequency $\hbar\omega/\left(SJ\right)=0.005953$
for the models with and without disorder. For the no-disorder model,
the skyrmion trajectory is a regular spiral with the radius increasing
with time. For the realistic model with disorder, the skyrmion is
cycling, too, but the center of the curvature is wondering and the
amplitude is not regularly increasing. Apparently, in the presence
of disorder the skyrmion is losing its energy into other types of
excitations, that is, disorder causes additional damping in the system.

Exactly at the AMC point, the SCR frequency is zero, and the skyrmion
motion cannot be excited by a harmonic perturbation. In this case,
one can apply a spin-current pulse to initiate the skyrmions' shift
$\mathbf{d}$, and after that the skyrmion will move ballistically,
as shown in Figs. \ref{Fig_Ballistic_motion} and \ref{Fig_Ballistic_motion_disorder}.
For the no-disorder model, ballistic motion is straight with reflections
from the boundaries. In the realistic model with disorder, ballistic
motion follows a random trajectory.

\section{Discussion}

We have shown analytically and numerically that the dynamics of a
skyrmion in a two-sublattice ferrimagnet differs significantly from
its dynamics in a ferromagnet. The differences arise from the possibility
of a specific deformation of the ferrimagnetic skyrmion that corresponds
to the separation of centers of skyrmions belonging to different sublattices.
Such a deformation of the ferrimagnetic skyrmions results in its finite
mass as compared to zero mass associated with the motion of the center
of topological charge of a skyrmion in a ferromagnet.

We have studied coupled Thiele equations for ferimagnetic sublattices
and obtained the dynamics of skyrmions in limiting cases. The finite
skyrmion mass generates its gyroscopic motion, which manifests as
a spin excitation specific to ferrimagnets -- skyrmion cyclotron
resonance (SCR)-- similar to electron cyclotron resonance (ECR) in
metals. The frequency of the SCR is given by a universal formula that
we derived analytically and tested in a numerical experiment on a
discrete spin lattice. Our numerical model is based on a Hamiltonian
tailored for parameters of the CoGd ferrimagnet, for which our predictions
can be tested by real experiments.

The frequency of the SCR goes to zero on approaching the angular momentum
compensation (AMC) point. In a TM/RE ferrimagnet, the AMC is achieved
at a certain concentration of the RE atoms. The hybridization of the
SCR with other ferrimagnetic modes dramatically changes the spectrum
of the excitations modes of the ferrimagnet in the presence of skyrmions.
With the skyrmion lattice in the background, the frequency of the
acoustic ferrimagnetic mode decreases at the AMC instead of going
up as it does in the absence of skyrmions.

When RE atoms are distributed randomly, the excitation modes are affected
by disorder. However, even in this case, the SCR is well defined and
has an amplitude comparable to that in a system with no disorder.
Similarly, the hybridization effects mentioned above remain pronounced
in the presence of disorder. They can be observed by exciting the
SCR with microwaves or by a spin current. Measurements of the SCR
must permit an unambiguous determination of the skyrmion mass, similar
to how the ECR in metals permits determination of the electron effective
mass.

We also studied ballistic and gyroscopic trajectories of skyrmions
in a ferrimagnet. Skyrmion cyclotron orbits must typically have a
small radius. For that reason, they may be difficult to visualize
in experiments, unless studied very close to the AMC, where their
radius diverges. In the presence of disorder, the cyclotron orbits
of skyrmions resemble the diffusive motion of electron cyclotron orbits
in metals.

We hope that this work inspires experiments on massive dynamics of
skyrmions in ferrimagnets, and the skyrmion cyclotron resonance in
particular. While our analytical and numerical methods have been developed
for a two-sublattice uniform ferrimagnetic film, we expect our results
for the dynamics of ferrimagnetic skyrmions and the excitation spectrum
of a ferrimagnet with skyrmions to be qualitatively valid for synthetic
ferrimagnets constructed of adjacent ferromagnetic layers \citep{Soum-Nat2017,Xia-PRB2021,Mallik-Nat2024}. 

\section*{Acknowledgments}

This work has been supported by Grants No. FA9550-24-1-0090 and FA9550-24-1-0290,
funded by the Air Force Office of Scientific Research.\\

\end{document}